\documentclass[sn-aps]{sn-jnl}% American Physical Society (APS) Reference Style
\jyear{2021}%

\theoremstyle{thmstyleone}%
\theoremstyle{thmstyletwo}%

\theoremstyle{thmstylethree}%

\usepackage{psfrag}
\usepackage{siunitx}
\usepackage{xcolor}

\usepackage{comment}
\usepackage{ulem}

\begin{document}

\title[Turbulent wake in fully-established transonic buffet]{Experimental investigation on the turbulent wake flow in fully-established transonic buffet conditions}


\author*[1]{\fnm{Christopher Julian} \sur{Schauerte}}\email{c.schauerte@aia.rwth-aachen.de}

\author[1]{\fnm{Anne-Marie} \sur{Schreyer}}\email{a.schreyer@aia.rwth-aachen.de}


\affil*[1]{\orgdiv{Institute of Aerodynamics}, \orgname{RWTH Aachen University}, \orgaddress{\street{Wüllnerstraße 5a}, \city{Aachen}, \postcode{52062}, \state{NRW}, \country{Germany}}}

%\affil[2]{\orgdiv{Department}, \orgname{Organization}, \orgaddress{\street{Street}, \city{City}, \postcode{10587}, \state{State}, \country{Country}}}
%
%\affil[3]{\orgdiv{Department}, \orgname{Organization}, \orgaddress{\street{Street}, \city{City}, \postcode{610101}, \state{State}, \country{Country}}}


\abstract{The transonic flight regime is often dominated by transonic buffet, %associated with 
a highly unsteady and complex shock-wave/boundary-layer interaction involving major parts of the flow field. % which is typically referred to as transonic buffet. 
The phenomenon is associated with a large-amplitude periodic %back and forth 
motion of the compression shock coupled with large-scale flow separation and intermittent re-attachment. Due to the resulting large-scale variation of the global flow topology, also %the development of 
the turbulent wake of the airfoil or wing  is severely affected, and so are any aerodynamic devices downstream on which the wake impinges. To analyze and understand the turbulent structures and dynamics of the wake, we performed a comprehensive experimental study %is performed in 
of the near wake %field 
of the supercritical OAT15A airfoil  in transonic buffet conditions at a chord Reynolds number of \num{2e6}. Velocity field measurements reveal severe global influences of the buffet mode %afftecting 
on both the surface-bound flow field on the suction side of the airfoil and the %near 
wake. The flow is %, and indicate an 
intermittently strongly separated, %nature 
with a significant momentum deficit that extends far into the wake. %near-wake field. 
The buffet motion induces severe disturbances and variations of the turbulent flow, as shown on the basis of %A detailed assessment of 
phase-averaged turbulent quantities in terms of Reynolds shear stress and RMS-values. % emphasizes severe disturbances induced by the buffet motion that persist up until one chord length downstream of the trailing edge. 
The spectral nature of downstream-convecting fluctuations and turbulent structures are analyzed using high-speed focusing schlieren sequences. %capture the spectral nature of downstream-convecting fluctuations and upstream traveling disturbances as well as give access to their spectral nature. 
Analyses of the power spectral density pertaining to the vortex shedding in the direct vicinity of the trailing edge indicate dominant frequencies one order of magnitude higher than those associated with shock buffet ($St_c=\mathcal{O}({1})$) vs. $St_c=\mathcal{O}({0.1})$). It is shown that the flapping motion of the shear layer is accompanied by the formation of a von Kármán-type vortex street of fluctuating strength. These wake structures and dynamics will impact any downstream aerodynamic devices affected by the wake. Our study therefore allows conclusions regarding the incoming flow of devices such as the tail plane.
}




\maketitle

\section{Introduction}
\label{sec:Intro}

% the intricacy of 
% inherently rooted in

% ---- Intro to transonic shock buffet
The transonic flight regime is often characterized by the occurrence of local supersonic regions on the suction side of airfoils and wings, which are terminated by a shock wave \citep{McDevitt1985}. For sufficiently strong interactions, shock-induced flow separation may span the entire region from the shock to the trailing edge \citep{Molton2013}. For certain combinations of Mach number, Reynolds number, and angle of attack, the transonic flow about the airfoil becomes unsteady, and %with 
the compression shock and separated boundary layer carry %ing 
out periodic oscillations \citep{McDevitt1985}. The resulting complex shock-wave/boundary-layer interaction involves major parts of the flow field \citep{Jacquin2009} and is typically referred to as transonic buffet.
% ---- Aerodynamic impact of buffet
The self-sustained dynamic back-and-forth motion of the shock wave is accompanied by strong variations of the shock strength, thus inciting a periodic thickening and shrinking of the separated shear layer \cite{Iovnovich_Raveh2012}. This self-excited periodic phenomenon is, inter alia, associated with a highly unsteady flow field \citep{Molton2013}, which %in turn 
results in low-frequency fluctuation of the integral aerodynamic quantities, i.e. the %namely 
lift and drag coefficients \citep{Crouch2008,Benoit_Legrain1987}. 

% ---- Structural coupling

For specific frequencies of this periodic change between a 
severely separated boundary layer and periods of mostly 
reattached flow, 
the aerodynamic phenomenon and the aircraft structure may interact. %be triggered. 
This mechanical coupling with the airframe can lead to a %detrimental 
resonance of the wing (the so-called buffeting), inciting a potentially deleterious effect on the structural integrity %of the aircraft
 \citep{Benoit_Legrain1987,Crouch2008}. 

% ----
%With an increasing demand for fast air travel, the overall efficiency  of modern civil aircraft becomes increasingly interesting. 
As the buffet onset %practically 
constrains the usable range of the flight envelope \citep{Benoit_Legrain1987, Molton2013}, its understanding is one of the major challenges in the design of modern civil aircraft \citep{Crouch2019,Jacquin2016}. %and a prerequisite to identify and potentially shift the limits of the flight envelope. 
The %large 
number of applications of supercritical airfoils increases with the increasing demand for fast air travel. Therefore, the overall efficiency  of modern civil aircraft becomes more and more relevant and %increasingly interesting ... 
the significance of a thorough understanding of the buffet phenomenon increases, as it is a prerequisite to identify and potentially shift the limits of the flight envelope. 

% ---
Several possible explanations for the underlying mechanisms %triggering 
of this oscillatory shock motion have been suggested. The well-established model proposed by \cite{Lee1990} and refined by \cite{Hartmann2013} %captures 
describes this %unsteady fluid 
motion as %based on 
a feedback loop. According to the model, two sequential phases of downstream and upstream propagation of disturbances form the recurring buffet cycle. The shock motion  initiates pressure waves that subsequently propagate downstream to the trailing edge. The interaction %Upon interacting 
with the latter emits disturbances that %are emitted and 
travel upstream along the airfoil surface. These disturbances provide the %amount of 
energy required to maintain the back and forth motion of the shock wave and thus close the feedback loop. Crouch et al. \cite{Crouch2008} linked the buffet onset with a global instability of the flow, and thus contributed significantly to the perception of the phenomenon and enhanced the prediction of the buffet onset. 

% ---- transition: research question --> wake
The majority of studies concerned with transonic buffet aim at contributing to a %are motivated by the prospect of 
profound understanding of the underlying mechanisms to mitigate buffet and finally prevent the structural response. %, i.e. buffeting. Despite seven decades of research on mechanisms that initiate and maintain this self-sustained unsteady shock-wave motion \citep{McDevitt1985, Lee1990}, however, several questions still remain unanswered \cite{Giannelis2017}. 
Substantially less attention has been dedicated to the evolution of the buffet-induced turbulent %near 
wake. %within one chord length of the trailing edge. The level of knowledge concerning its potential impact on the performance aerodynamic devices such as control surfaces of modern aircraft designs remains vague and almost unexplored. Under the aspect of 
Regarding the practical implementation in civil aircraft, however, also the influence of the wake 
on 
aerodynamic devices downstream of the wing, such as tail plane control surfaces, is crucial, as it may affect their performance. % of downstream aerodynamic devices such as tail plane control surfaces.

%Several numerical and experimental studies on t
The formation of wakes downstream of airfoils has been studied %conducted 
for either low-Reynolds number flows \cite{Gunasekaran2018}, non-buffet conditions \cite{Epstein1988, AlshabuOlivier2008}, or mostly %substantially 
laminar flow configurations \cite{Zauner2020}. 
%
%The %particular 
%combination of fully-established shock buffet in turbulent flow at high-Reynolds number has been addressed by 
Only Szubert et al.\,\cite{Szubert2015} addressed fully-established shock buffet in turbulent flow at high-Reynolds number and %, who 
replicated the conditions identified relevant for pronounced buffet by Jacquin et al. \cite{Jacquin2009} in terms of Mach number, angle of attack, and chord-based Reynolds number.
%Some publications strongly suggest a coupling between t

The suction-side-bound oscillatory flow mode of the airfoil %\textit{(Was wurde in den genannten Studien untersucht? 3D Flügel oder 2D Profil?)}} 
and the complex dynamics inherent %ly ascribed 
to the turbulent wake appear to be coupled \cite{Szubert2015,Zauner2020}. %These studies thus promote a
A similarly complex interplay between the two domains %, which 
was also observed %earlier  in a similar manner 
for other transonic flows that comprise pronounced vortex streets, e.g. double-wedge profiles with strong separation \cite{Franke1989}. 

For flow configurations with %comprising 
large wake fluctuations in the context of unsteady shock motion, Lee et al. \cite{Lee1994} studied the formation of upstream traveling waves. %was studied systematically by , and 
These waves were advanced as an essential element to predict the periodic shock oscillation \cite{Lee1990}. %As these waves were interpreted by 
Tijdeman \cite{Tijdeman1977} interpreted these waves as an immediate consequence to changes in the flow conditions at the trailing edge of the airfoil and denoted them %they were denoted 
as Kutta waves. 

%In their study on 
For laminar buffet conditions, Zauner and Sandham \cite{Zauner2020} identified a complex interplay between Kelvin-Helmholtz instabilities in the surface-bound flow towards the rear part of the airfoil,  the formation of a von Karman vortex street aft of the trailing edge, and subsequently shed pressure waves traveling upstream towards to the shock wave. Based on a modal analysis of an established buffet flow using proper orthogonal decomposition, Szubert et al. \cite{Szubert2015} showed %were able to show 
an %mutual 
interaction between the vortex shedding mode and the buffet mode. %This study points out that e
Events of massive flow separation – not necessarily driven by developed buffet – may thus give rise to von Kármán-like vortex shedding in the vicinity of the trailing edge and beyond in the near wake \cite{Szubert2015}. Pressure sensors placed along the airfoil upper surface and in the near wake %were able to 
captured strong secondary fluctuations (in addition to the primary buffet oscillation)  over a broad range of chordwise stations. These fluctuations were attributed to a von Kármán mode and persisted throughout the entire near-wake domain downstream of the trailing edge \cite{Szubert2015}. %Based on p
Power spectra %l density analyses performed on 
of the wall-pressure signals showed the characteristic low-frequency ($St_{c}=0.075$) footprint of shock buffet. 

% ---- 
%Taking account of 
For a more global, integral perspective of an entire aircraft, %configuration 
it is %of great interest to not only grasp the onset of buffet dynamics which will in turn lead to an enhanced exploitation of the flight envelope but also 
necessary to also quantify the dynamics that the buffet phenomenon imposes on the %surrounding 
aircraft structure and control surfaces via the wake of the wing or airfoil. %in the range of the wake. 
The quantitative importance of the wake unsteadiness to the %integral 
aerodynamic performance of a civil transonic aircraft has yet to be determined.
As a first step, the present work %is concerned with a detailed assessment of 
assesses the %transition 
region around %between 
the airfoil rear section and the near wake in detail for %domain in 
fully-established buffet conditions, since fully-established buffet represents the most severe dynamic conditions. %The quantitative importance of the wake unsteadiness to the %integral 
%aerodynamic performance of a civil transonic aircraft has yet to be determined

We characterize the temporal and spatial structure of the separated boundary layer and near wake %is characterized 
with particular focus on the variation over the buffet cycle. A careful assessment of the wake structure and involved dynamics is %therefore considered 
the essential groundwork for understanding the %future exploration of an even more 
complex interaction of a fully-established buffet configuration with a tail plane. To the authors’ knowledge, this is the first experimental investigation to focus explicitly to the evolution of the buffet-dominated periodic fluctuation of the near wake field by means of optical flow measurement techniques.

In Sec.\,\ref{sec:ExpSetup}, we introduce the experimental methods and setup, followed by a detailed discussion of the turbulent flow field in Sec.\,\ref{sec:Results}. We present our conclusions in Sec.\,\ref{sec:Conc}.

\section{Experimental methods and 
setup}
\label{sec:ExpSetup}

We briefly introduce %A brief introduction to 
the experimental facility and investigated airfoil models, before presenting %is provided before 
an %concise 
overview of the applied measurements arrangement. % is presented. 
The section is complemented with %some 
fundamental considerations of vector field post-processing that form the basis for subsequent sections.

	\subsection{Wind-tunnel facility and model}
	
Our experiments %of this study 
were conducted in the closed-circuit Trisonic Wind Tunnel at the Institute of Aerodynamics, RWTH Aachen University. The Trisonic Wind Tunnel is an intermittently-operated vacuum-type indraft facility. Based on interchangeable test sections and adjustable nozzle geometries, it can provide Mach numbers between $0.3 < \mathrm{M}_{\infty} < 4.0$. %, thus covering sub-sonic, transonic, as well as supersonic flow conditions. 
Continuously run screw-type compressors evacuate a vacuum chamber with a total volume of $380 \, m^3$ downstream of the test section and transport the air through a dryer bed before it enters a large $180 \,m^3$ air reservoir at ambient conditions. Turnaround times of approximately 7 minutes are achieved. The relative humidity of the air is kept below 4\,\% to preclude condensation effects and falsifications of the measured shock location \citep{Binion1988}. %Upon initialization of a 
A measurement cycle is initialized by the actuation of a fast-acting valve that allows %is actuated,  allowing 
the high and low pressure sections to retrieve equilibrium conditions, thus drafting the measurement medium %to be sucked 
from the balloon through the test section into the vacuum tanks. In each measurement cycle, %an effective acquisition phase of 
stable flow conditions are obtained for 2 to 3 seconds, depending on the configured Mach number. The obtainable Reynolds number is a function of the selected Mach number and the ambient conditions of the resting, pre-dried air in the reservoir. The turbulence intensity based on the freestream velocity fluctuation of the wind tunnel is below 1\,\%. 

For the present study, measurements were conducted at
$\mathrm{M}_{\infty}=0.72$, which is %was varied between 0.68 and 0.80, 
equivalent to a free-stream velocity of $u_{\infty}=235$~\si{\meter/\second}. %freestream velocities \ams{between} $u_{\infty}=224$~\si{\meter/\second} and $u_{\infty}=257$ \si{\meter/\second}. 
For the investigated airfoil, a %configuration, a %s with different chord lengths, 
chord-based Reynolds number of $Re_c = 2 \cdot 10^6$ is obtained. These %measurement 
conditions require %Measurements of the present campaign are performed at Mach numbers in the high subsonic regime, equivalent to 
the transonic operation mode. The transonic test section features a rectangular cross section of 400 mm x 400 mm and a total length of 1,410 mm.
\begin{figure}[hbt!]
\centering
\includegraphics[height=.25\textheight]{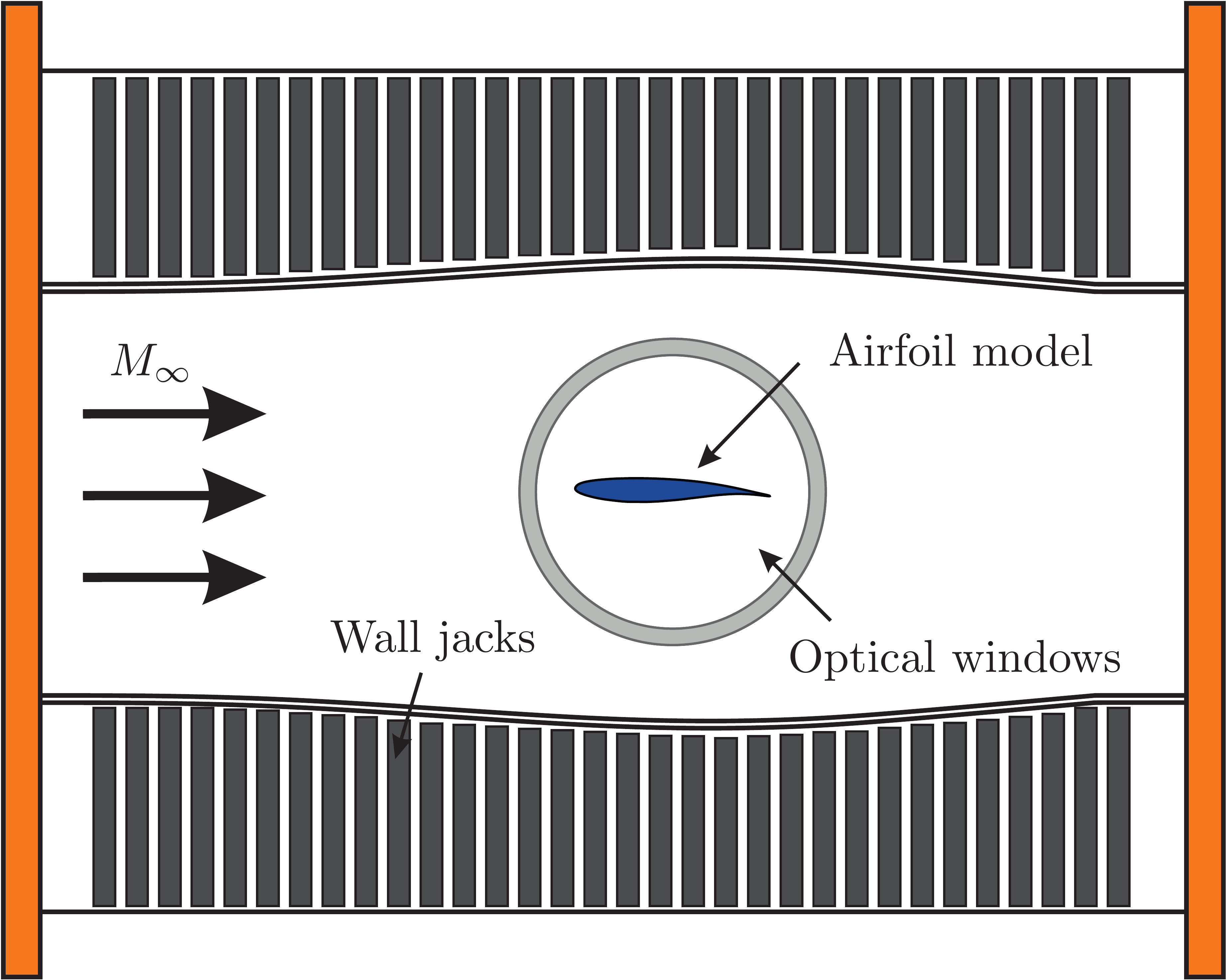} \qquad
\caption{Illustration of the transonic test section with adjustable walls} %and optical access windows}
\label{fig:wind_tunnel}
\end{figure}
%
%The transonic test section 
It is equipped with flexible upper and lower walls to compensate the %account for 
influence of solid walls confining the flow and to emulate conditions in terms of streamline contouring observed in the equivalent free, unconfined flow around the same airfoil geometry \citep{McDevitt1982}. %Fig. \ref{fig:wind_tunnel} \ams{shows a} schematic representation of the airfoil installed in the optically accessible test section. %is provided in Fig. \ref{fig:wind_tunnel}. %The obtainable Reynolds number is a function of the selected Mach number and the ambient conditions of the resting, predried air in the reservoir. 
\begin{comment}
For the present study, measurements were conducted at
$\mathrm{M}_{\infty}=0.72$, which is %was varied between 0.68 and 0.80, 
equivalent to a free-stream velocity of $u_{\infty}=235$~\si{\meter/\second}. %freestream velocities \ams{between} $u_{\infty}=224$~\si{\meter/\second} and $u_{\infty}=257$ \si{\meter/\second}. 
For the investigated airfoil configuration, a %s with different chord lengths, 
chord-based Reynolds number of $Re_c = 2 \cdot 10^6$ is obtained. 
\end{comment}

%	\subsection{Airfoil model}
	
We studied buffet on the supercritical OAT15A airfoil.  The two-dimensional wind tunnel model has a relative thickness of 12.3 \% and features a total span of 399 mm, which is equivalent to the width of the test section. A chord length of $c=150\,\text{mm}$ was chosen, giving an aspect ratio of approx. 2.66. The upper and lower half of the model were manufactured from stainless steel, along with a trailing edge made of metal particle-filled epoxy resin. The center of rotation for the angle of attack is located at about mid-chord of the airfoil.

Fig. \ref{fig:wind_tunnel} shows a schematic representation of the airfoil installed in the optically accessible test section.

	\subsection{Measurement arrangement}
	
\label{sec:FocSchlieren}

Flow visualizations are obtained with a focusing schlieren setup, where extended grids illuminated by correspondingly large illuminated surfaces replace the point-shaped light sources used in classical schlieren arrangements \cite{Schardin1942,Settles2001}. A focusing schlieren system inspired by Weinstein's approach \cite{Weinstein1993} has been developed by the authors \citep{Schauerte_Schreyer2018} %(see Fig.\,\ref{fig:SchlierenSetup}) 
to explore flow configurations in the transonic and supersonic %Mach number 
flow regime. %, which are frequently accompanied by shock-wave/boundary layer interaction and large-scale separation \cite{Delery1986}. 

This system overcomes some of the limitations of classical schlieren configurations, in particular the close to infinite depth of focus that results in strong line-of-sight integration. We require the ability to focus on narrow slices of the flow to be able to analyze our highly unsteady and partially three-dimensional flow of interest. 

%To temporally resolve the oscillatory motion of the shock, the schlieren visualizations were recorded with a Photron SA-5 CMOS high-speed camera at a frame rate of 20 kHz and with a short exposure of \SI{1.9}{\micro\second}. The flow was illuminated by a 3x3 array of 15W high-power LEDs together with a constant-current source with an electric output adjustable between 50 W and 120 W. %that allows for continuous variable operation between 50 W and 120 W electric output. 
The flow was illuminated by a 3x3 array of 15W high-power LEDs together with a constant-current source with an electric output adjustable between 50 W and 120 W.
The source grid consists of CNC-machined alternating clear and opaque lines, %whereas 
the cutoff grid is %designed as 
an exact photographic negative of the source grid. %to ensure ideal match. 
A large 470 mm Fresnel lens, placed slightly ahead of the source grid, %fulfills several purposes, i.e. it 
captures major portions of the light intensity, illuminates the source grid evenly bright by reshaping the light cone, and redirects the light beam through the test section past the model and onto the schlieren lens. The illuminated measurement area around the spanwise centerplane of the airfoil has a diameter of 280 mm. However, the outer edges are subject to vignetting due to the light beam passing %being passed 
through two consecutive circular windows. An effective field of view (FOV) without vignetting of $d_{FS} \approx 220 \textrm{mm}$ was achieved, using a magnification $M_{FS}$ of 0.2 between the two grids in conjunction with the relay optics and a 35 mm camera lens.

To temporally resolve the oscillatory motion of the shock, the schlieren visualizations were recorded with a Photron SA-5 CMOS high-speed camera at a frame rate of 20 kHz and with a short exposure of \SI{1.9}{\micro\second}. 
For the high-speed recording, a FOV of \SI{180}{\milli\meter} width and  \SI{130}{\milli\meter} height is projected on the camera sensor, yielding an effective resolution of 4.05 px/mm.

The focusing schlieren setup was operated %in principal 
%in two configurations, (a) 
with a horizontally aligned cutoff grid %orientation 
to capture %primarily 
vertical density gradients. %, and (b) with vertical %the 
%grid lines %being vertical which allows 
%to resolve mainly horizontal gradients.
%With regard to the present test series, t
From these measurements, we extracted the temporal development of the SWBLI and the near wake, the shock position and amplitude, as well as the shock oscillation frequencies. 

To acquire detailed velocity data, we applied particle image velocimetry (PIV). %system. 
Planar measurements are performed in a streamwise/wall-normal plane along the centerline of the airfoil model. %is selected in the present application. 

\begin{figure}[hbt!]
\centering
\includegraphics[height=.25\textheight]{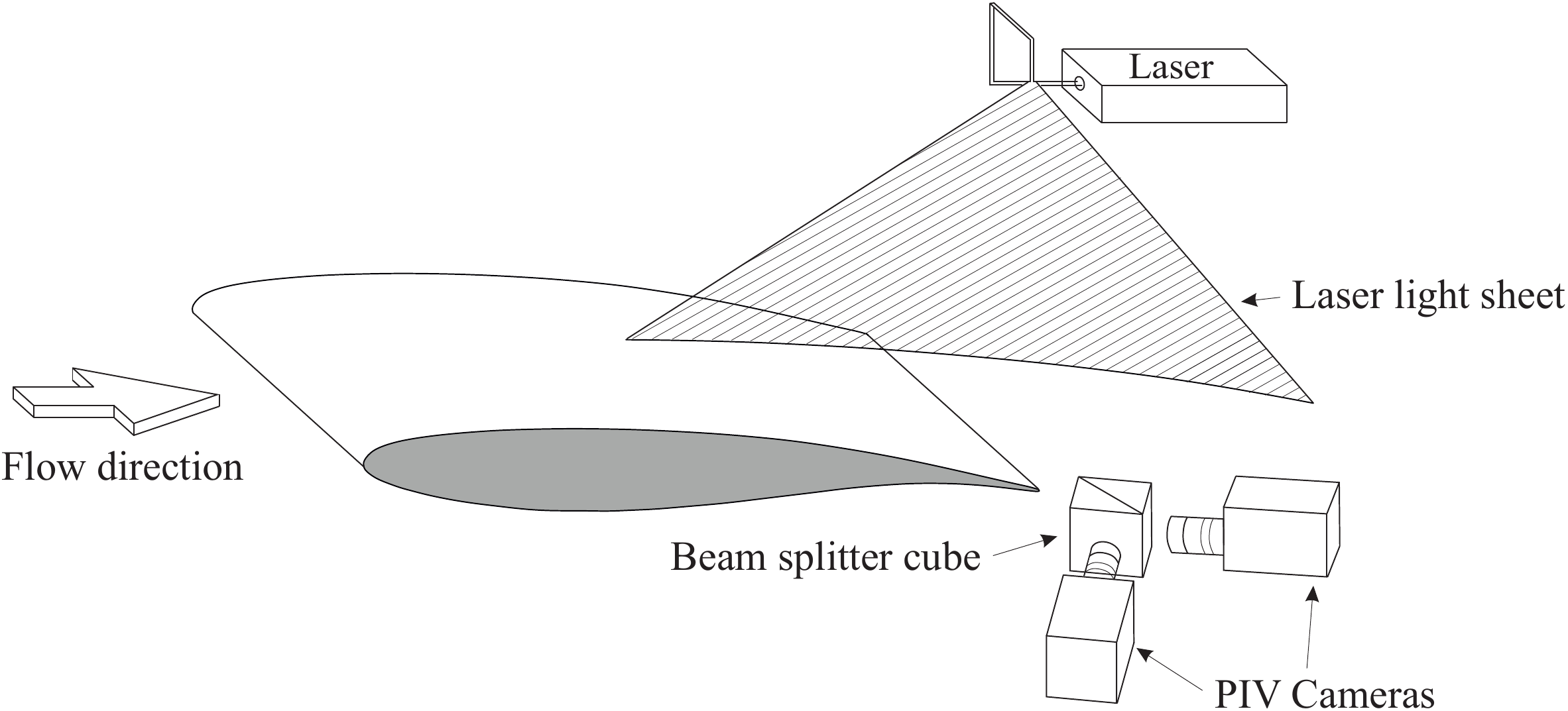} \qquad
\caption{Overview of the measurement location captured by the PIV setup}
\label{fig:Model_measurement_setup}
\end{figure}

A Litron Nano PIV double pulsed Nd:YLF laser, operating at a wavelength of 527 nm, produces a light sheet of 1.5 mm thickness. The large laser-pulse energy of 200 mJ/pulse with a pulse duration of 4 ns ensures a high signal-to-noise ratio. The flow is seeded with di-ethyl-hexyl-sebacate (DEHS) tracer droplets with a mean diameter of 1 $\mu m$. Particle images are captured using two identical FlowSense EO 11M CCD cameras with a resolution of 11 megapixels, which corresponds %is equivalent 
to a sensor resolution of 4,008 pixels x 2,672 pixels at a pixel size of $9\,\mu m$. Both cameras are focused on the same field of view (FoV) to increase the acquisition rate to 10 Hz. The FoV covers both the vicinity of the airfoil trailing edge and the near wake. It is \SI{145}{\milli\meter} wide and \SI{80}{\milli\meter} high, thus spanning a streamwise domain from 0.6\,c to 1.7\,c. %in chordwise direction. 
Based on the magnification of 4.04, a spatial resolution of \SI{27.56}{px \per \milli\meter} is obtained. With a final interrogation-window size of 16\,px x 16\,px at an overlap of 50\,\%, the resulting velocity maps feature a vector spacing of 0.29 mm. Relevant parameters of the focusing schlieren and PIV tests are summarized in table \ref{tab:ExperimentalParameters}.

\begin{table}
\caption{Overview of experimental PIV and FS parameters.}
\label{tab:ExperimentalParameters}
\centering
\begin{tabular}{lllllll}
\hline
%Transition & & & \multicolumn{2}{c}{}\\\cline{1-3}
\multicolumn{7}{l}{Particle image velocimetry setup\hspace{2.1cm} ~ Focusing Schlieren setup}\\ 
%\hline
\cline{1-3} \cline{4-7}
Sensor resolution & 4008 x 2672 & px x px & & Sensor resolution & 704 x 520 & px x px \\
%\cline{1-6}
FOV width & 150 & mm & & FOV width & 180 & mm \\
%\cline{1-6}
FOV height & 80 & mm & & FOV height & 150 & mm \\
%\cline{1-6}
Acquisition rate & 10 & Hz & & Acquisition rate & 20,000 & Hz \\
%\cline{1-6}
Exposure & 11 & \SI{}{\micro\second} & & Shutter speed & 1/525,000 & s \\
%\cline{1-6}
Spatial resolution & 27.56 & px/mm & & Spatial resolution & 4.05 & px/mm \\
\hline
\end{tabular}
\end{table}

\subsection{Identification of buffet phases}
	\label{sec:IdentificationBuffetPhases}
\begin{figure}[hbt!]
\centering
\includegraphics[width=.98\textwidth]{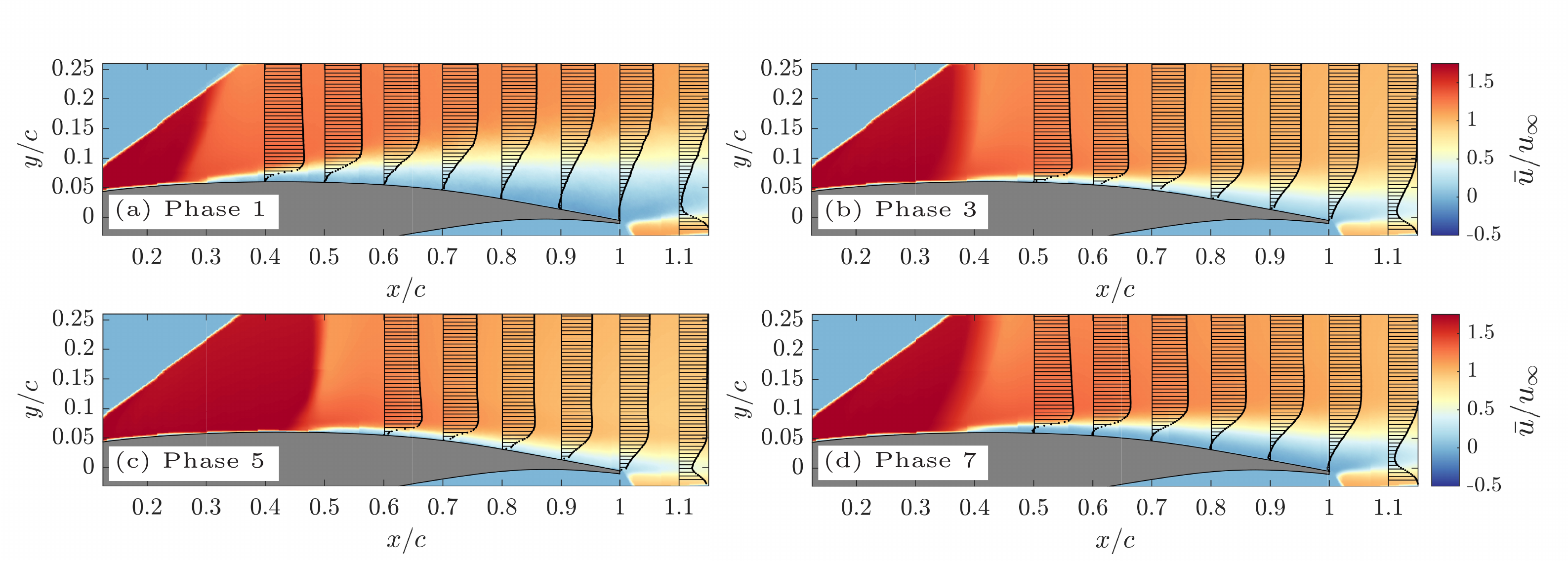}
\caption{Phase-averaged streamwise velocity maps and profiles for characteristic phases of the buffet cycle.}
\label{fig:u_phases_profiles_upstreamFOV}
\end{figure}

Preceding analyses of the %exploration of a 
buffet %-dominated 
flow around the OAT15A airfoil (see \cite{Schauerte_Schreyer2022a}) %for more details) 
with a field of view covering the entire suction side from 0.1\,c to 1.1\,c.  %of the airfoil 
revealed large-scale global modifications of the flow topology throughout the buffet cycle (see also Fig.\,\ref{fig:u_phases_profiles_upstreamFOV}). %ascribed to the buffet mode. 

Since the shock motion covers %To account for the extensive 
%back and forth %motion by \ams{along} 
almost $20\,\%$ of the chord and %this motion 
is coupled with intermittent flow separation and reattachment, %\ams{we} deemed 
statistics %a statistical evaluation %based on 
for discrete phases of the buffet cycle are more meaningful than %instead of 
a global mean. The contour maps in Fig. \ref{fig:u_phases_profiles_upstreamFOV} show the shock's most upstream location (a), an intermittent chord wise postion during downstream (b), and upstream (d) motion, as well as the shock most downstream location (c).

\begin{figure}[hbt!]
\centering
\includegraphics[width=0.98\textwidth]{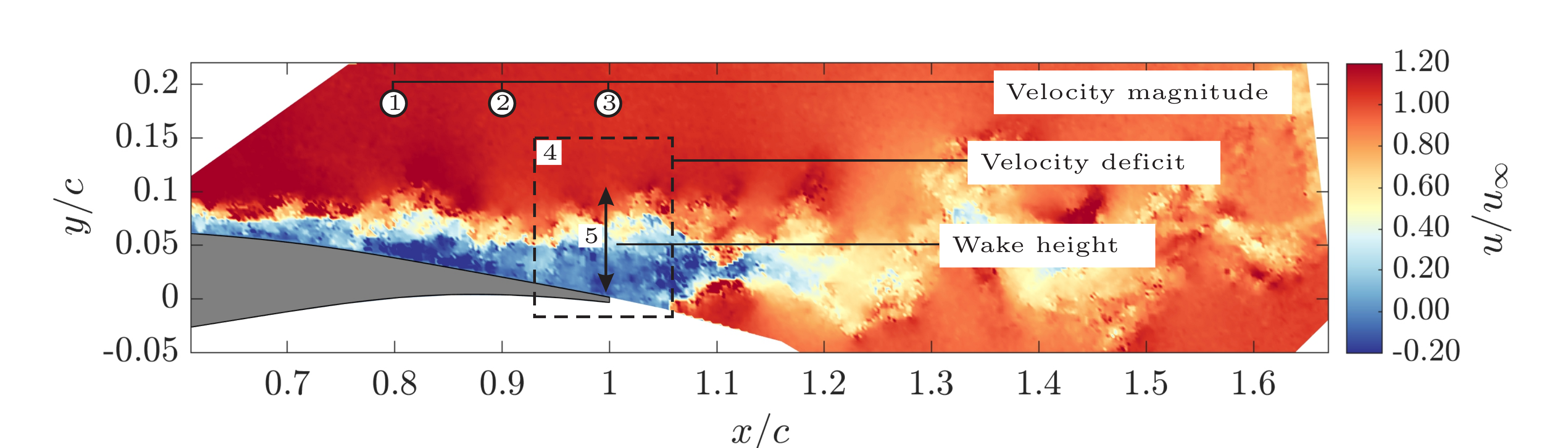} \qquad
\caption{Phase assignment for %performed on 
instantaneous velocity fields in the wake} %field of view}
\label{fig:PhaseAveragingWake}
\end{figure}

Individual buffet phases can %could 
be unambiguously assessed based on the instantaneous shock location, the shape of the shock front, and the shock inclination near the airfoil contour for each uncorrelated snapshot \cite{Schauerte_Schreyer2022a}. Based on a sufficiently large overlap between the %two 
fields of view in the present and the earlier \cite{Schauerte_Schreyer2022a} measurement campaign, the latter quantities were %then 
applied to the rear portion of the airfoil and the near wake, % domain, 
yet excluding the actual shock wave. The procedure and evaluated parameters are clarified in Fig.\,\ref{fig:PhaseAveragingWake}.

First, the integral velocity magnitude within the rectangular analysis window in Fig.  \ref{fig:PhaseAveragingWake} (4) is evaluated in terms of the percentage of vectors that exceed or stay below a certain threshold ($u\leq 0.25\,u_{\infty}$). Threshold values were previously identified to uniquely describe the chordwise location of the shock and thus allow the %enable a 
definition of characteristic phases within the buffet cycle. %containers. That 
This procedure allows %for a first pre-assignment of 
to pre-assign the %uncorrelated 
snapshots to the respective buffet phases. This criterion was complemented by considering the height of the low-momentum bubble (see Fig.  \ref{fig:PhaseAveragingWake} (5)). However, due to similar velocity defects for phases of final upstream motion and incipient downstream motion, some ambiguity remains in this first classification. % of instantaneous velocity maps. 

Second, the streamwise velocity magnitude %was extracted 
at three chordwise stations (location (1), (2), and (3); Fig. \ref{fig:PhaseAveragingWake}) %for each snapshot. These %obtained 
%values 
were %then 
compared with reference values from the predefined phase assignment (see Fig. \ref{fig:u_phases_profiles_upstreamFOV}) for each snapshot. 

An analysis %detailed assessment 
of the thus finally obtained phase assignment in the %upstream 
field of view covering the entire airfoil suction side (studied in \cite{Schauerte_Schreyer2022a}) revealed that each buffet phase induces characteristic velocity defects. %are induced \ams{each} %, depending on the 
%buffet phase
These deficits are modulated by the direction of motion and strength of the shock wave. This third criterion is %can be considered 
independent of the velocity deficit in the strongly separated region and can %is able to 
eliminate the last ambiguity in the phase assignment. 

Finally, as a consistency check, the velocity profiles at the chordwise stations between $x/c=0.8$ to $x/c=1.1$ of the present case and the case studied in \cite{Schauerte_Schreyer2022a} were %superimposed and 
examined for %their 
congruence. The results showed excellent agreement (less than $0.6\%$ deviation), which proofs that we can %therefore 
successfully identify the respective %between the results demonstrated the successful reconstruction of 
buffet phase of a snapshot without directly knowing %knowledge of 
the shock position.

\section{Results and discussion}
\label{sec:Results}

Since the flow varies considerably throughout a buffet cycle, and these variations have a strong impact on the wake of the airfoil, we begin this section with a short discussion of %n introductory assessment of 
the flow topology on the airfoil suction side for %surface in the present 
fully-estasblished buffet conditions. We then briefly discuss the rear section of the airfoil and near-wake region ($0.6\leq x/c\leq 1.7$) on the basis of %These results are followed by 
ensemble-averaged velocity maps. % covering the near wake before we elucidate further details of the 
Subsequently, we discuss the wake structure in detail, using phase-distinguished representations of instantaneous and averaged velocity fields in the streamwise and vertical directions. We also provide a %as well as 
discussion of turbulent quantities and the vorticity, as well as a spectral analysis of the vortex shedding at the trailing edge. %complement the results section.

\subsection{Shock-buffet induced variation of the flow topology} % and assessment of the %established 
%buffet cycle}
\label{sec:UpstreamFoV}
	
\begin{figure}[hbt!]
\centering
\includegraphics[width=.7\textwidth]{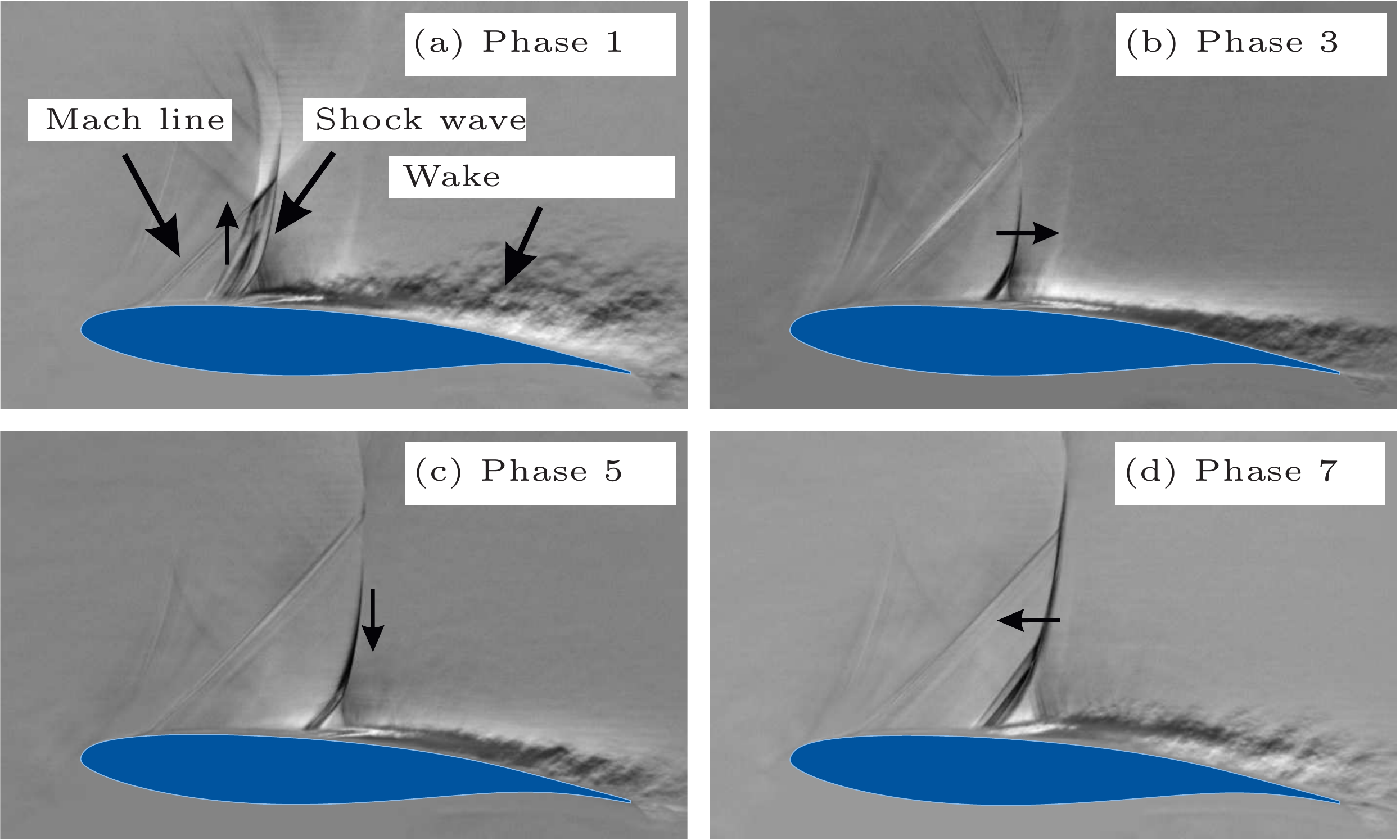}
\caption{Focusing schlieren snapshots capturing the global variation of the flow topology over the full buffet cycle.}
\label{fig:FocusingSchlierenPhases}
\end{figure}

In our previous study \cite{Schauerte_Schreyer2022a,Schauerte_Schreyer2022b}, we acquired high-speed focusing schlieren sequences %were acquired to capture 
capturing the temporal evolution of the flow topology, in particular %involving 
the low-frequency shock oscillation. An overview of selected phases of the buffet cycle is given in Fig.\,\ref{fig:FocusingSchlierenPhases}, which emphasizes the huge periodic variation of the flow, synchronized with a back and forth motion of the shock wave that covers approx. $20\,\%$ of the airfoil chord. The images indicate a strongly turbulent wake with numerous turbulent eddies of varying size. %showing evidence an intensely eddying turbulent wake following a shock-induced flow separation and emphasizing the huge periodic variation of the flow synchronized with a back and forth motion of the shock wave at about $20\,\%$ of the airfoil chord. 

We identified eight characteristic phases throughout a buffet cycle, primarily determined by the respective shock location and direction of shock motion (either upstream or downstream).
Phase 1 (Fig. \ref{fig:FocusingSchlierenPhases} (a)) corresponds to the most upstream shock location close to the reversal point, for which the separated boundary layer thickens strongly and the wake is of large surface-normal extent. %a massive thickening of the separated boundary layer and a large wake are observed. 
The downstream motion of the shock wave %(top row of Fig. \ref{fig:FocusingSchlierenPhases}) 
is accompanied by a continuous straightening of the shock and a successive reduction in separation \cite{Schauerte_Schreyer2022a, Schauerte_Schreyer2022b}, which can be seen in phase 3 according to Fig. \ref{fig:FocusingSchlierenPhases} (b). %collapse of the separated boundary layer. 
The minimum extent of the separation region is reached in phase 5 (see Fig.\,\ref{fig:FocusingSchlierenPhases} (c)), around the shock's most downstream location.
While the shock wave is %substantially 
oblique in phase 1 (shock is at an angle of about $70^\circ$ to the horizontal line), it is %has taken a shape 
almost normal to the surface in phase 5 (Fig.\,\ref{fig:FocusingSchlierenPhases} (c)). At the onset of the shock's upstream motion in phase 7 (see Fig.\,\ref{fig:FocusingSchlierenPhases} (d)), the shock starts to tilt back and the separation region increases. %ed boundary layer 
%grows in the vertical direction. 
The distinct lambda shock pattern is characteristic of %represents a characteristic feature in 
this type of flow and is discernible in all snapshots, however the vertical triple point location changes considerably %is subject to sizeable changes again strictly coupled 
with the current chordwise location of the shock wave and its direction of motion.

To %adequately 
account for the large-scale variation of the global flow topology, we carried out a phase-averaged description of the velocity fields from PIV according to the phases discussed in the previous paragraph and following the procedure outlined in Sec.\,\ref{sec:IdentificationBuffetPhases}. %that is associated with the fully-established buffet condition,  and to complement the focusing schlieren assessment presented in Fig. \ref{fig:FocusingSchlierenPhases}, a phase-averaged description of the velocity field is required. %In total, eight characteristic phases, primarily determined by the respective shock location and direction of motion (either upstream or downstream) were defined. 
%In total, 5,300 snapshots capturing the airfoil suction side were processed to obtain t
Fig.\,\ref{fig:u_phases_profiles_upstreamFOV} shows the resulting phase-averaged streamwise velocity contours and %corresponding 
profiles for the phases corresponding to the Schlieren visualizations shown in Fig.\,\ref{fig:FocusingSchlierenPhases}. %as presented in Fig. \ref{fig:u_phases_profiles_upstreamFOV}.
%	
%\begin{figure}[hbt!]
%\centering
%\input{Images/BuffetCyclePhases/u_phases_profiles_upstreamFOV.tex}
%\includegraphics[width=.98\textwidth]{Images/BuffetCyclePhases/u_phases_profiles_upstreamFOV.eps}
%\caption{Phase-averaged streamwise velocity maps and profiles for characteristic phases of the buffet cycle}
%\label{fig:u_phases_profiles_upstreamFOV}
%\end{figure}
%
We present a brief summary of the strongly varying flow topology and behavior throughout a buffet cycle here, since it lays the basis the subsequent exploration of the wake properties. An in-depth discussion of this data is provided in \cite{Schauerte_Schreyer2022a} and \cite{Schauerte_Schreyer2022b}. %For the present case, this data set serves as an essential groundwork to understand the subsequent exploration of the wake properties as it allows for an elucidation of the highly fluctuant nature of the buffet mode. Therefore, a brief outline of the previous work will be presented here.

The first phase includes all instantaneous flow representations with far upstream shock locations. It is characterized by massive flow separation (see Fig. \ref{fig:u_phases_profiles_upstreamFOV} (a)). %, which starts %forming 
%slightly downstream of the shock foot. % and grows strongly for $x/c\geq 0.5$. 
The severity of the flow separation is further substantiated in the great velocity deficit %we observe 
in the near-wake velocity profile at $x/c=1.1$. Phase 2 (not shown) represents the phase of incipient downstream motion of the shock wave. Compared with phase 1, the shock line is less oblique. %and has moved from $x/c \approx 0.23$ to $x/c \approx 0.28$ in the phase-averaged representation (see Fig. \ref{fig:u_phases_profiles_upstreamFOV} (b)). 
This trend is seen to continue in phase 3 (Fig. \ref{fig:u_phases_profiles_upstreamFOV} (b)), in which the %Fig.\,\ref{fig:u_phases_profiles_upstreamFOV} (c)), 
vertical extent of the wake and its related momentum deficit have reduced by almost $75\,\%$ compared with phase 1. %, and about $50\,\%$ when compared with phase 2. The streamwise velocity component is consistently positive across the vertical domain, and free stream conditions are reached at $y/c \approx 0.07$. The shock wave occupies locations between $x/c=0.35$ and $x/c=0.38$. 
The continued downstream displacement of the shock wave from $0.35\,c$ to $0.40\,c$ between phases 3 and 4 is accompanied by a further contraction of the turbulent wake, most apparent in %the much 
fuller velocity profiles close to the wall (not shown here; see \cite{Schauerte_Schreyer2022a,Schauerte_Schreyer2022b}). %at $x/c=1.0$ (Fig.\,\ref{fig:u_phases_profiles_upstreamFOV} (d)). 
Phase 5 (see Fig.\,\ref{fig:u_phases_profiles_upstreamFOV} (c)) denotes the most downstream shock location $x/c \approx 0.48$. The shape of the shock line is now exactly normal, and the $\lambda$-region has reached a minimal extent. The flow is completely attached all along the suction side, and the wake influence is only noticeable in the immediate vicinity of the airfoil contour. 
%The phase of the buffet cycle captured in p
Phases 6 (not shown) and 7 (Fig.\,\ref{fig:u_phases_profiles_upstreamFOV} (d)) capture the %defines 
incipient upstream motion of the shock. The maximum upstream shock velocity is not yet reached, however, both wake size and the velocity deficit close to the wall are greater than in the corresponding equivalent shock location in phases 3 and 4 during the shock downstream motion. %While t
The velocity profiles of phase 7 are %was 
significantly less full than for %the fully-recovered 
phase 5, and the wake increases. %phase 7 is characterized by a largely increased wake (see Fig.\,\ref{fig:u_phases_profiles_upstreamFOV} (g)). 
The $\overline{u}$ velocities therefore contain negative values all the way until the TE. The distinct obliqueness of the shock line and the large $\lambda$-structure are both finely reproduced in the PIV data. The final phase of the shock upstream motion is represented by phase 8 (not shown), in which the identified low-momentum wake protrudes even farther into outer flow field than in the previous phase, and the induced momentum deficit is significantly more severe. The deficit even results in a small region of reverse flow close to the airfoil surface with a vertical extent of $y/c \approx 0.025$ above the trailing edge, comparable to phase 1 (Fig.\,\ref{fig:u_phases_profiles_upstreamFOV} (a)).

\subsection{Evolution of the shock-induced wake throughout the buffet cycle}
\label{subsec:BuffetCycle_wake}

\begin{figure}[hbt!]
\centering
\includegraphics[width=0.7\textwidth,clip=true]{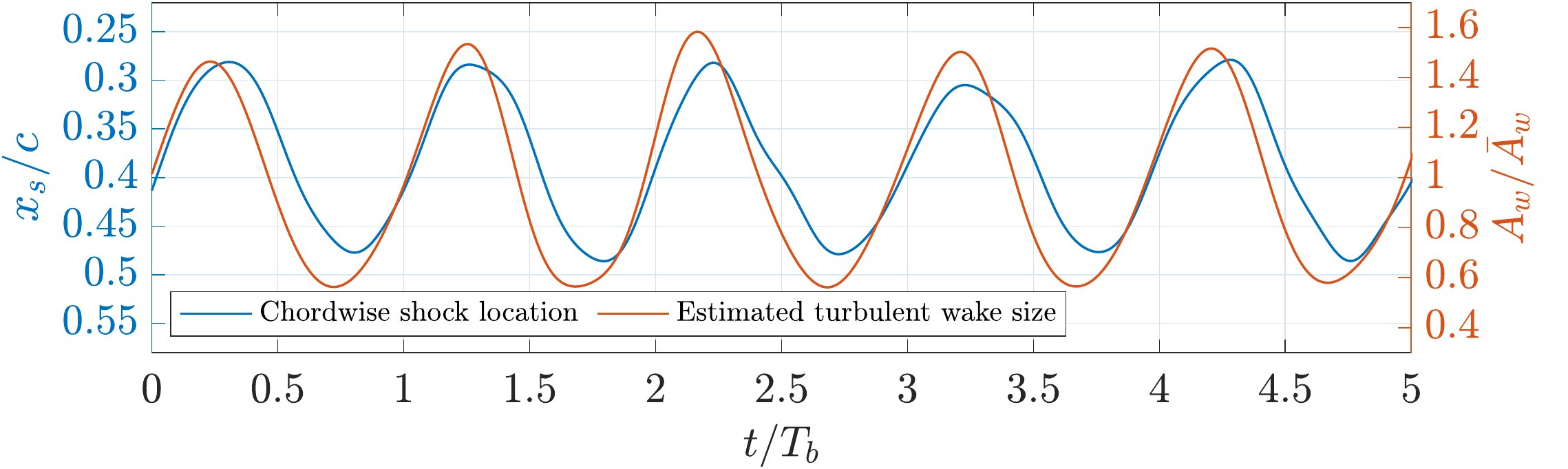} 
\caption{Temporal development of upstream and downstream shock motion (in blue) in comparison with the evolution of the highly turbulent wake region (in red).} 
\label{fig:PhaseShockPosWakeSize}
\end{figure}

The extent of the turbulent wake varies strongly throughout the buffet cycle (compare Fig.\,\ref{fig:FocusingSchlierenPhases}). Its extent is coupled with the direction of motion of the shock. Since the shock motion is periodic, we expected a phase coupling of the wake size with the shock position.  To verify this assumption, we estimated the relative size of the turbulent shear layer and wake along the chord length throughout the buffet cycle.  A gradient-based detection was applied to the time-resolved schlieren sequences to extract the respective edges of the shear layer. Then, the area enclosed by these edges was calculated and used as an estimate of the turbulent wake area $A_w$. Overall, we evaluated 60 complete buffet cycles, i.e. over 10,000 images. The time history of the shock position $x_s$ (in relation to the chord length) and separated wake area (normalized with its temporal mean extent $\overline{A}_w$) are shown in Fig. \ref{fig:PhaseShockPosWakeSize}. 

The quantities seem to be phase-locked and thus oscillate at an identical frequency, but at an almost constant small phase shift. The wake area precedes the shock motion by approximately 0.3491 rad at its most upstream, and by 0.5585 rad at its most downstream location. The wake size reaches its minimum extent shortly before the shock occupies its most downstream location. The wake size already begins to increase before the shock passes its reversal point, and reaches its maximum towards the final phase of the shock upstream movement. Overall, %A further feature to note is that 
the wake area %undergoes a variation in size by which it 
almost doubles its size within a full buffet cycle and performs a pronounced, recurrent flapping motion.

		\subsection{Mean velocity maps in the near wake}

\begin{figure}[hbt!]
\centering
\includegraphics[width=.90\textwidth]{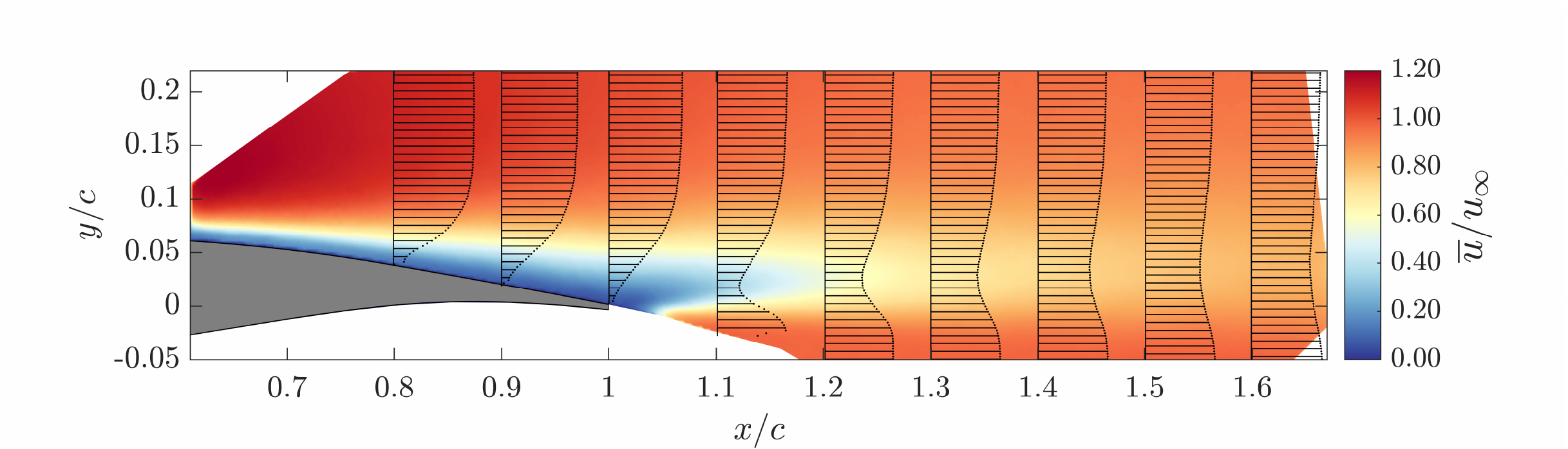}
\caption{Ensemble-averaged streamwise velocity map and profiles at characteristic chordwise locations in the near wake %field 
for fully-established buffet %condition 
at M=0.72 and AoA=5 deg.}
\label{fig:u_mean_global}
\end{figure}

\begin{figure}[hbt!]
\centering
\includegraphics[width=.90\textwidth]{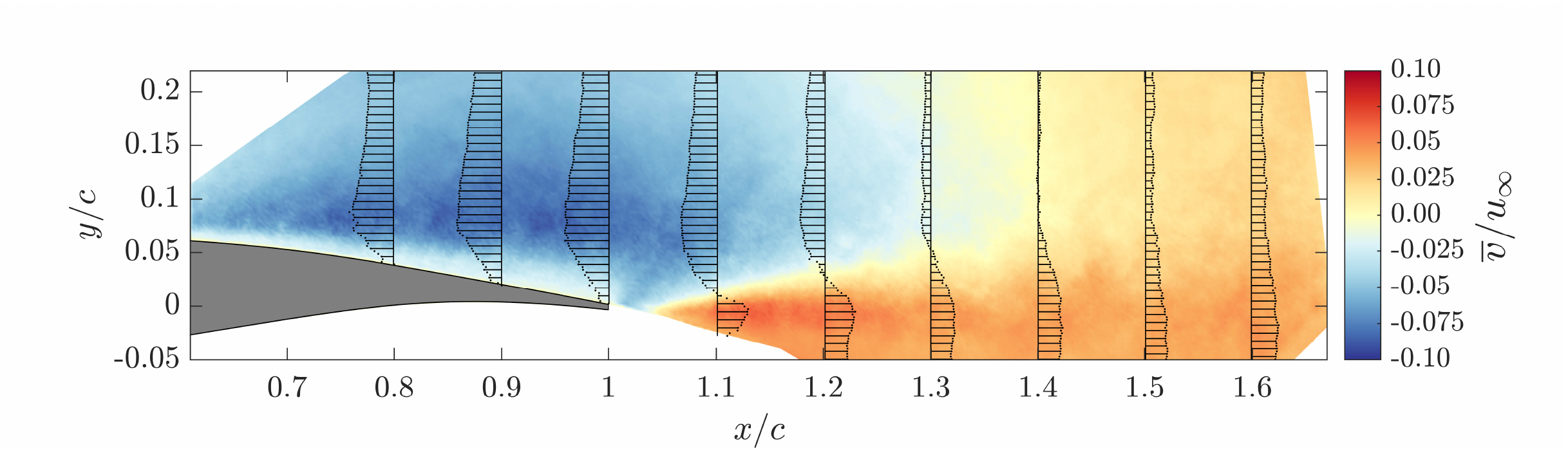}
\caption{Ensemble-averaged vertical velocity map and profiles at characteristic chordwise locations in the near wake %field 
for fully-established buffet %condition 
at M=0.72 and AoA=5 deg.}
\label{fig:v_mean_global}
\end{figure}

Before discussing the strong periodic variation of the wake flow field, we present %The 
ensemble-averaged velocity maps of the streamwise %component in Fig. \ref{fig:u_mean_global} 
and vertical velocity components in Figs.\,\ref{fig:u_mean_global} and \ref{fig:v_mean_global}, respectively. This %condense the highly-unsteady buffet flow configuration to a 
global representation approximates the temporally-averaged flow physics and %thus 
allows to assess the integral impact of the buffet state. 

The mean streamwise contour map (Fig.\,\ref{fig:u_mean_global}) indicates a large region of %is indicative of an 
intermittently %strongly 
separated flow: the momentum deficit %in close proximity of 
near the trailing edge %closely 
approaches $\overline{u}/u_{\infty} \approx 0$. As will be shown later in the context of Fig. \ref{fig:u_phases_profiles}, this region is %mainly 
a result of the vigorous shock dynamics, %inherent to buffet, and 
in particular phases of upstream %traveling 
shock motion, for which reverse flow extends over large parts of the rear of the airfoil up %wake and is seen to exist 
until the trailing edge.  %Even though t
The greatest velocity deficit occurs close to the surface and in %is limited to 
the inner wake core region (visible as blue %here indicated by the bluish 
streak %along the rear section of the airfoil and downstream of the trailing edge and denoted by 
with $u \leq 0.25\,u_{\infty}$ in Fig. \ref{fig:u_mean_global}). The low-momentum dent increases with downstream development. Furthermore, %, a successive widening of the low-momentum dent is apparent with growing $x/c$. From a global perspecitve, we further observe a slight upwards displacement of the wake center with increasing $x/c$.
the wake center shifts slightly upwards.

The %distribution of the 
wall-normal velocity field is structured into %exhibits a clear structure in the ensemble-averaged representation. From Fig. \ref{fig:v_mean_global}, we extract 
three characteristic subdomains (see Fig.\,\ref{fig:v_mean_global}): a pronounced downwash region %tendency 
along the aft section of the airfoil, an %clear 
upwash region downstream of the trailing edge, which emanates from the pressure side, and a region of almost vanishing $v$-component further downstream of the airfoil in the upper half of the wake. These %tendencies 
regions are %nicely 
captured by the velocity profiles at chordwise stations $x/c=1.1$ to $x/c=1.4$. %While a
At $0.1\,c$ aft of the trailing edge, a sharp distinction between up- and downwash is %still 
dominant. The upwash from the pressure side seems to gradually overcompensate the downwash component, yielding a close-to-zero vertical flow component in the upper half of the captured domain at $x/c=1.4$.
		
		\subsection{Modification of the wake velocity field by buffet} %mode}
Despite the considerable level of turbulent perturbations %comprised 
in instantaneous flow representations,  they %still 
convey important features characterizing the nature of the dynamic flow field
 that are precluded in phase- %averaged 
or ensemble-averaged velocity contours. These features include %Several aspects of the nature of the wake such as 
turbulent structures, shedding of vortices, and the overall extent %penetration 
of the turbulent wake. %\ams{into the overall flow field (?)}. %to name a few are thus unveiled, promoting a deeper insight in established the buffet-dominated flow.

\begin{figure}[hbt!]
\centering
\includegraphics[width=.98\textwidth]{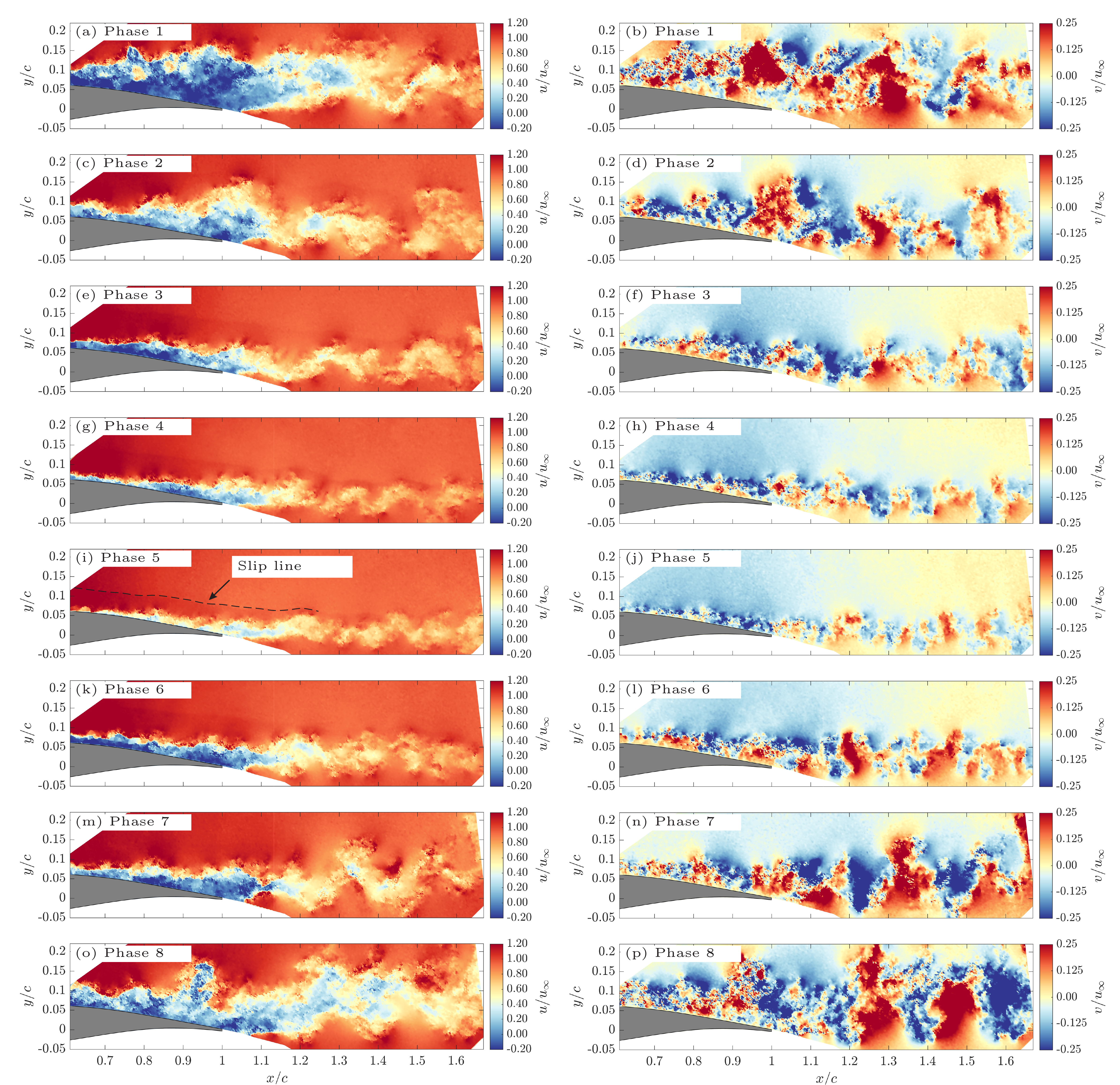}
\caption{Instantaneous streamwise (left column) and vertical (right column) velocity maps for characteristic phases of the buffet cycle}
\label{fig:u_phases_instantaneous}
\end{figure}

The instantaneous %representation of the 
vertical velocity contours %capture the %severely disturbed 
show a turbulent wake of large extent in all phases of far upstream shock locations, namely phases 1, 2, 7, and 8 (see Figs. \ref{fig:u_phases_instantaneous} (b), (d), (n), (p)). In comparison, %to that, 
the flow topology gradually recovers when %with 
the shock %wave progressing 
progresses towards its most downstream location (phases 3, 4, and 5 in Figs.\,\ref{fig:u_phases_instantaneous} (e), (h), and (k). Turbulent %fluctuations 
structures are confined to a narrow band, as most clearly visible in phase 5 (Fig. \ref{fig:u_phases_instantaneous} (j)). %, turbulent fluctuations are confined to a narrow band in the horizontal extension of the trailing edge whereas the mangitude of external flow velocity is evenly distributed.

%One sees that t
The slip line emanating from the triple point location of the lambda shock (marked with a dashed line in Fig.\,\ref{fig:u_phases_instantaneous} (i)) is nicely captured in the streamwise component (left column of Fig. \ref{fig:u_phases_instantaneous}), and most clearly reproduced in %less separated 
phases 4, 5, and 6 (Fig. \ref{fig:u_phases_instantaneous} (g), (i),  and (k)) with smaller separation regions. %This is due to t
The large separation in the remaining phases %which 
widens the lambda %shock 
structure and pushes the triple point upwards.

Phases of pronounced momentum deficit and large-scale separation %strongly enlarged boundary layer size 
(phases 1, 2, 7, and 8) show concomitant intense perturbations in the wall-normal velocity. These manifest themselves %itself 
in a regular pattern of alternating %spikes 
velocity regions of opposite sign, forming interlaced structures with %individual 
positive (upwards) contributions from the pressure side and negative (downwards) contributions from the suction side. %Both 

The size and intensity of these velocity perturbations vary considerably in sync with the buffet cycle. %unsteadiness. 
Coherent patches of %either 
positive or negative wall-normal velocity fluctuation are largest when the shock is far upstream, and shrink %gradually for increasingly 
with downstream shock displacement. %Under the aspect of downstream influence, it should be noted that t
The fluctuation magnitude of $\pm 0.25 u_{\infty}$ remains almost constant between $x/c=1.0$ and $x/c=1.7$, which %thus 
suggests a relevant influence of this phenomenon even farther downstream.

\subsection{Phase-averaged organization of the wake velocity field}
\begin{figure}[hbt!]
\centering
\includegraphics[width=.98\textwidth]{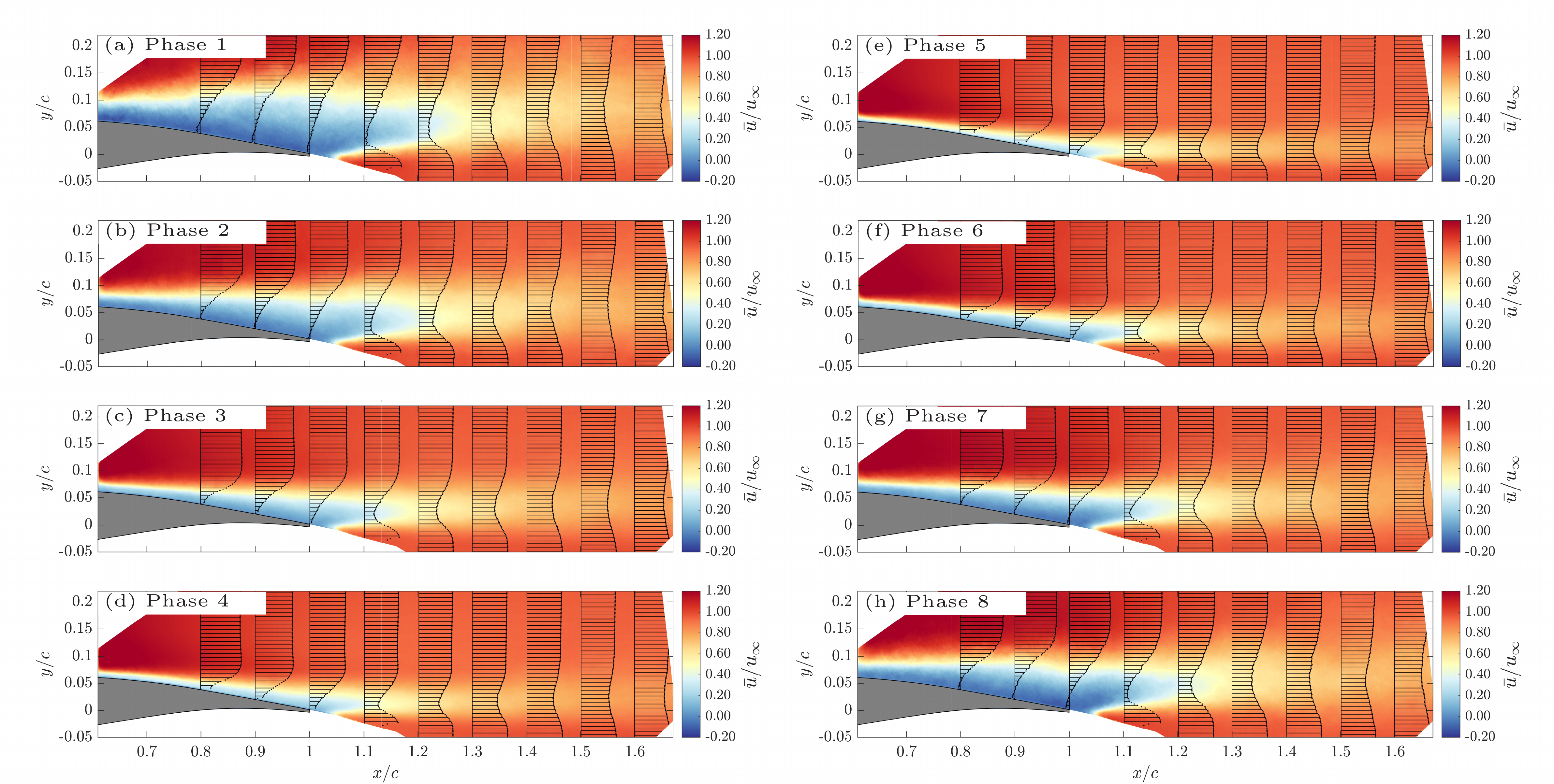}
\caption{Phase-averaged streamwise velocity maps and profiles for characteristic phases of the buffet cycle}
\label{fig:u_phases_profiles}
\end{figure}

\begin{figure}[hbt!]
\centering
\includegraphics[width=.98\textwidth]{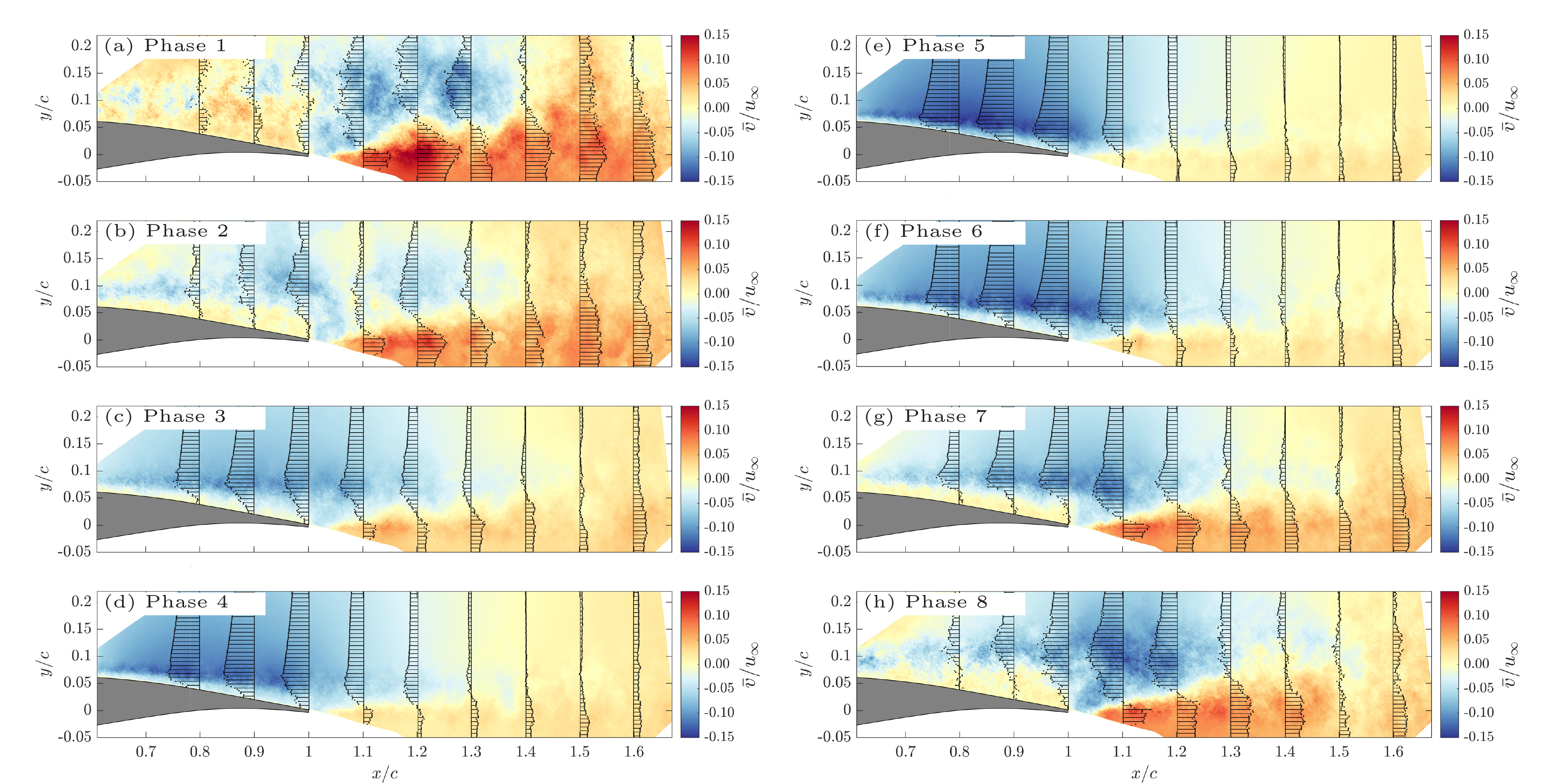}
\caption{Phase-averaged vertical velocity maps and profiles for characteristic phases of the buffet cycle}
\label{fig:v_phases_profiles}
\end{figure}

The large-scale variation of the global flow topology associated with the buffet mode in the present 2D configuration %was elucidated in Fig. \ref{fig:u_phases_profiles_upstreamFOV}. 
has been discussed in Sec.\,\ref{sec:UpstreamFoV}. In good agreement with these prior findings, the streamwise velocity contours in Fig. \ref{fig:u_phases_instantaneous} clearly indicate %d that 
the same buffet-governed variation %continues to exist 
across the near wake domain. % and involves large part of the flow field downstream of the trailing edge. 
To allow for a conclusive statistical evaluation of the velocity field and the inherent turbulent quantities, we provide a phase-related description of the wake velocity field analoguous to the discussion of the flow along the airfoil suction side following the procedure outlined in Sec.\,\ref{sec:IdentificationBuffetPhases}. %fully congruent with the discussion of the airfoil suction side (see Fig. \ref{fig:u_phases_profiles_upstreamFOV}) is considered beneficial. 
Such representions are provided for the streamwise and wall-normal components in Figs. \ref{fig:u_phases_profiles} and %for the wall-normal component in Fig. 
\ref{fig:v_phases_profiles}, respectively.
		
The contour maps %reproduced 
in Fig. \ref{fig:u_phases_profiles} show that the streamwise velocity maps vary periodically %undergo a cyclic variation synchronized 
with the buffet cycle. The low-velocity wake core fluctuates strongly, which is predominantly %with a strong fluctuation of the low-velocity wake core, mostly 
dictated by the respective location and direction of motion of the shock wave \cite{Schauerte_Schreyer2022a,Schauerte_Schreyer2022b}. The buffet mechanism leads to %is predominantly defined by 
a pronounced variation of the vertical and streamwise extent of the velocity-deficit region, which is associated with an intense flapping motion of the separated shear layer \cite{Schauerte_Schreyer2022a,Schauerte_Schreyer2022b}. 

%Whereas 
In the streamwise velocity magnitude, the greatest variation over the buffet cycle %in the streamwise velocity magnitude pertain to 
occurs with respect to its distribution the vertical direction (compare most extreme phases according to Figs.\,\ref{fig:u_phases_profiles} (a) and (e)). In the wall-normal component, dominant changes  are most evident in the streamwise direction (see Fig.\,\ref{fig:v_phases_profiles}) and are attributed to the interplay between up- and downwash from both sides of the airfoil downstream of the trailing edge. %These results are displayed in Fig. \ref{fig:v_phases_profiles}.

For the most upstream shock location (phase 1, see Fig. \ref{fig:u_phases_profiles} (a)), the severe flow separation %as well as the massive 
and thickening of the separated shear layer %associated with \ams{the} most upstream shock location (phase 1, see Fig. \ref{fig:u_phases_profiles} (a)) 
overcompensate the %natural 
curvature-related downwash %tendency 
in the rear part of the airfoil. Consequentially, a region of %and results in an overall 
mild upwash results %in the region 
between $x/c=0.6$ and $x/c=1.0$. 

Immediately downstream of the trailing edge, %upon coalescence of 
the flow fields from the upper and lower sides coalesce, and %flow fields, 
the pressure side induces an %substantial 
upwards deflection with $v/u_{\infty} \approx 0.1$ adjacent to a %bubble 
region of downwards flow above the trailing edge. This patch of negative vertical velocity persists until the chordwise station $x/c \approx 1.4$, beyond which the transverse component $v$ is %flow continues with 
consistently positive. %values $v$. 
In phases with upstream shock locations, i.e. %either 
incipient downstream motion (phase 2) or final upstream motion (phase 8), the downwash along the airfoil suction side is still deficient and less pronounced compared to phase of widely attached flow (phases 4, 5, and 6).

%The organization of the wall-normal velocity maps as for magnitude and spatial distribution in phases 3 and 7 are reminiscent of the ensemble-averaged field that was presented in Fig. \ref{fig:v_mean_global}, which reiterates the intermittent nature of this configuration.
Regarding the magnitude and spatial distribution, the organization of the wall-normal velocity maps in phases 3 and 7 are rather similar to the ensemble-averaged field shown in Fig.\,\ref{fig:v_mean_global}. This observation emphasizes the intermittent nature of this configuration.

\subsection{Evolution of turbulent quantities}
\begin{figure}[hbt!]
\centering
\includegraphics[width=.98\textwidth]{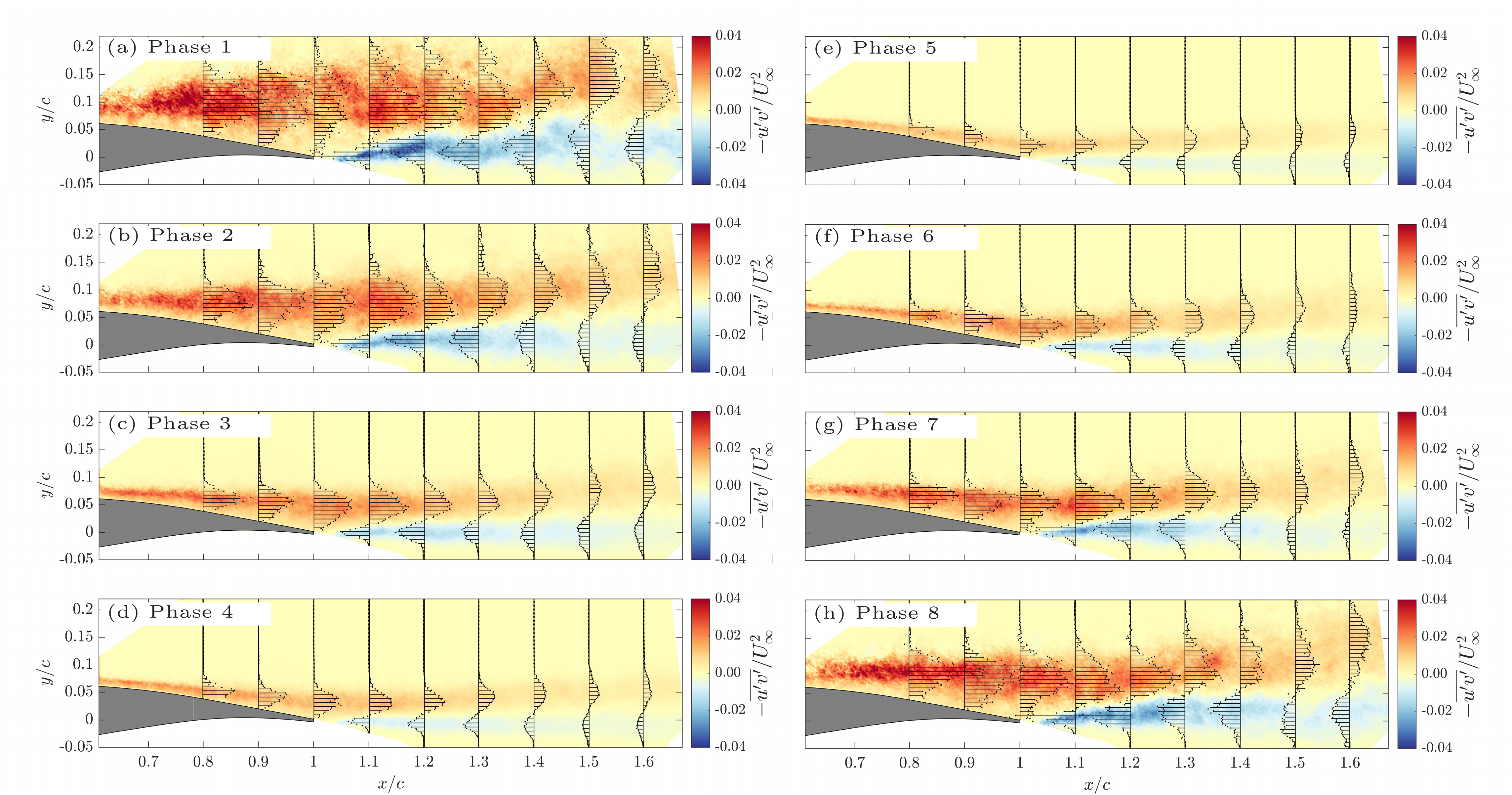}
\caption{%Overview of the 
Reynolds shear stress $Re_{uv}$  and its evolution during the buffet cycle.}
\label{fig:ReynoldsPlots_uv}
\end{figure}

\begin{figure}[hbt!]
\centering
\includegraphics[width=.98\textwidth]{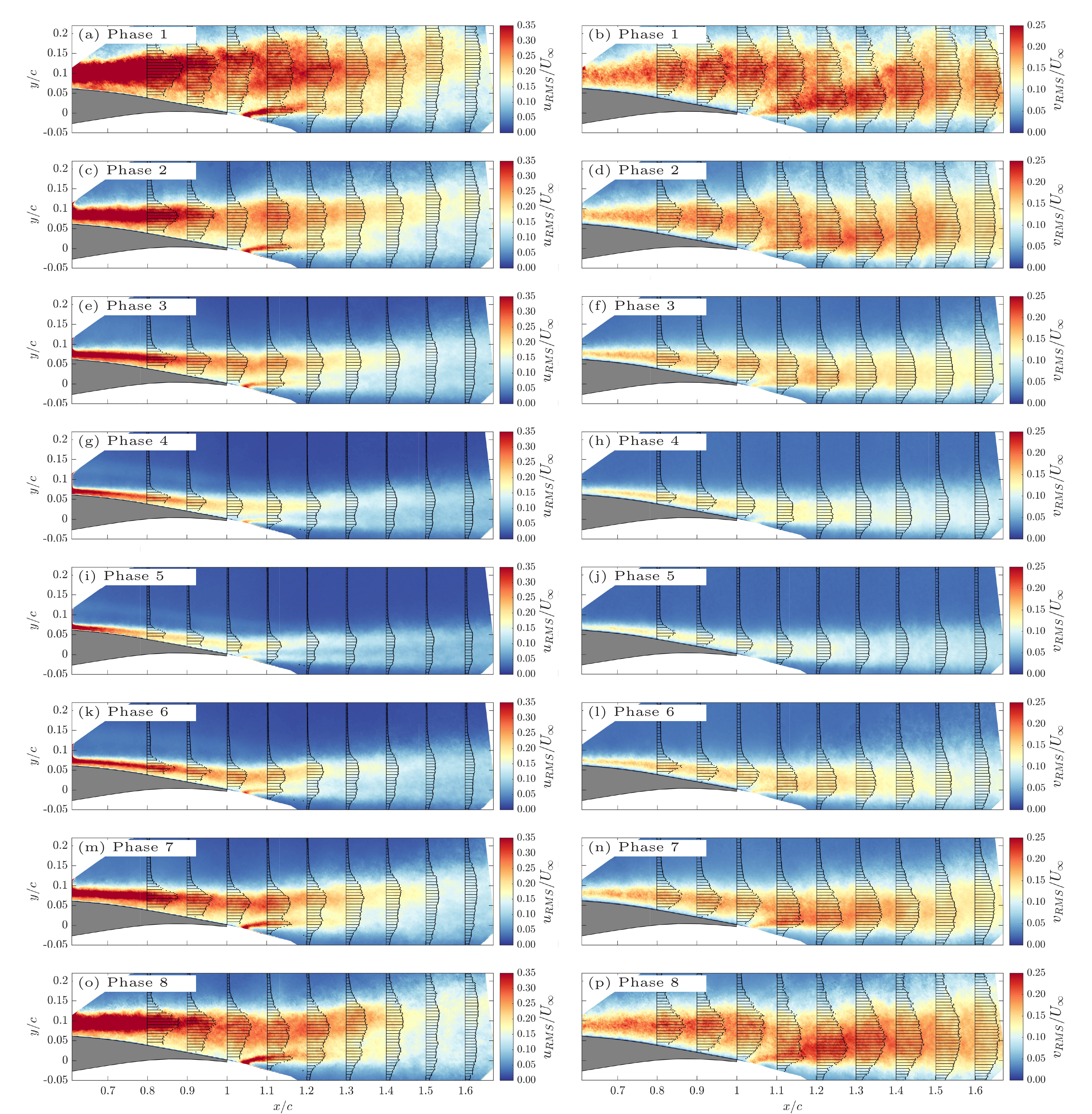}
\caption{Overview of the turbulent quantities $u_{rms}$ (right column) and $v_{rms}$ (left column).}
\label{fig:ReynoldsPlots_uu_vv}
\end{figure}

Considering the strong variations of the turbulent wake across the different buffet phases, %\ams{discussed in the previous sections,} 
a more detailed %elucidation 
analysis of the buffet-%ascribed 
related perturbations is required.
%Taking account of the previously discussed exemplary instantaneous velocity maps in $u$ and $v$ across the different buffet phases in the context of Fig. \ref{fig:u_phases_instantaneous},  a more detailed elucidation of the buffet-ascribed perturbations is required. Owing to the huge variance throughout the buffet period and to further quantify their spatial structure, t
The streamwise and transverse turbulent quantities $u_{rms}$ and $v_{rms}$ as well as the Reynolds shear stresses $Re_{uv}$ were computed for each phase individually (see Figs.\,\ref{fig:ReynoldsPlots_uv} and \ref{fig:ReynoldsPlots_uu_vv}) to assess their huge variations throughout the buffet cycle and further quantify their spatial structure. % and are reported in Figs.  \ref{fig:ReynoldsPlots_uv} and \ref{fig:ReynoldsPlots_uu_vv}.

%\begin{equation}
%Re_{uv} = \left(\frac{\sum\limits_{i=1}^n -u'v'}{n}\right)/u_{\infty}^2
%\end{equation}
%
%\begin{equation}
%u_{RMS}/u_{\infty} = Re_{uu} = \left(\sqrt{\frac{\sum\limits_{i=1}^n u'u'}{n}}\right)/u_{\infty}
%\end{equation}
%
%\begin{equation}
%v_{RMS}/u_{\infty} = Re_{vv} = \left(\sqrt{\frac{\sum\limits_{i=1}^n v'v'}{n}}\right)/u_{\infty}
%\end{equation}

The Reynolds shear stress contours in Fig.\,\ref{fig:ReynoldsPlots_uv} are consistent with our %are in accordance 
with prior findings \cite{Schauerte_Schreyer2022a} and agree with our observations above: intense turbulent mixing coincides with the incidents of large-scale flow separation, %are governed by intense turbulent mixing, 
as is most clearly %represented by 
visible in phases 1 and 8 (Figs.\,\ref{fig:u_phases_instantaneous} (b) and (p)). %subplots of phases (b) and (p) of  Fig.  \ref{fig:u_phases_instantaneous}. 

These effects are reflected in the shear-stress %contributions 
distributions of the same phases phases 1 (see Fig. \ref{fig:ReynoldsPlots_uv} (a)) and 8 (see Fig. \ref{fig:ReynoldsPlots_uv} (h)), whose respective maxima are %maximum is 
located far above the TE at $y/c = 0.1$ (phase 1) and $y/c = 0.075$ (phase 8), and the wake influence is still noticeable until $y/c = 0.2$ in both phases. %of the buffet cycle. 
Greatest maximum shear stress values of approx. $-\overline{u'v'}/u_{\infty}^2=0.04$ occur in %are attributed to 
phase 1, which confirms the qualitative observations of the instantaneous vector field in Figs.\,\ref{fig:u_phases_instantaneous} (a) and (b). %In contrast
As expected, the weakest shear stress of approx. $-\overline{u'v'}/u_{\infty}^2=0.015$ occurs in phase 5, %expectedly shows  
with a vertical peak location of $y/c = 0.025.$ This immense variation within the buffet cycle further illustrates that the wake carries out a strong flapping motion, as discussed in the context of Fig. \ref{fig:PhaseShockPosWakeSize} in Sec.\,\ref{subsec:BuffetCycle_wake} on the basis of the time history of the wake area estimate. % of the wake.

%Both quantities 
The streamwise and wall-normal turbulence components $u_{rms}$ and $v_{rms}$ show %a basically 
qualitatively similar distributions, %and magnitudes, % whereas the 
although some distinct features are visible in the shapes of their profiles (see Fig.\,\ref{fig:ReynoldsPlots_uu_vv} left and right columns, respectively). %exhibit characteristic features. 
Both quantities seem to emerge from two major source regions: %which are first, 
%an intense turbulent production in 
from the suction-side shear layer downstream of the shock wave %, and second, turbulence production bound to 
and from the lower side of the trailing edge. These %individual 
contributions are %well-
reflected in the respective profiles near the airfoil trailing edge. %provided in Fig. \ref{fig:ReynoldsPlots_uu_vv}. 
At $x/c=1.1$, as the %two flow fields 
flow of the suction and pressure sides coalesce for phases of far upstream shock locations (phases 1 and 8, see Figs.\,\ref{fig:ReynoldsPlots_uu_vv} (a), (b), (o), and (p)), the effect is most distinct: both $u_{rms}$ and $v_{rms}$ profiles exhibit two separate peak contributions (at $y/c \approx 0.1$ and $y/c \approx 0.0$), between which the turbulence intensity reduces. %Despite the qualitatively similar behavior, 
The distinctness of the peaks is much stronger for the streamwise $u_{rms}$ component. 
%One remarkable feature is the clear tendency that t
%The streamwise %fluctuation component
$u_{rms}$ is much more intense %concentrated 
along the two %individual 
wake main axes from the airfoil suction and pressure sides, whereas the wall-normal component is of more even magnitude across the width of the wake. This effect results %thus resulting 
in slightly M-shaped profiles for $u_{rms}$, e.g. at $x/c \geq 1.2$. 

%While f
For $u_{rms}$, %values, 
the contribution from the suction %generally 
dominates: the %, showing 
overall levels and extent in the $y$-direction at the location of the trailing edge are greater than %compared to 
the minor contribution from the pressure side. The vertical fluctuations (right column of Fig. \ref{fig:ReynoldsPlots_uu_vv}) are equally strong on either side. Furthermore, the wake contribution from the pressure side to the $u_{rms}$ component decays much faster than the contribution to $v_{rms}$ (at approx.\,$x/c\leq1.5$ in phase 1, whereas the contribution in $v_{rms}$ has not yet decayed at $x/c=1.7$).
	
\subsection{Evolution of vortical structures in the wake}

%Contours maps of instantaneous vorticity for three characteristic buffet phases are provided in Fig. \ref{fig:Vorticity_Phases_1_5_8}. 

We explore the shedding of vortices, expressed by the vorticity, in detail, since the intermittently strongly separated flow field %configuration observed 
in transonic buffet is %was 
associated with the formation of vortex streets \cite{Zauner2020}. %, the shedding of vorticity is explored in detail. 
The vorticity field was computed on the basis of the streamwise and wall-normal components of instantaneous velocity fields %contours in the streamwise and wall-normal direction were exploited to compute the vorticity field 
according to eq. \ref{eg:vorticity}.

\begin{equation}
\omega_{z} = \frac{\partial v }{\partial x} - \frac{\partial u }{\partial y}
\label{eg:vorticity} 
\end{equation}

\begin{figure}[hbt!]
\centering
\includegraphics[width=.98\textwidth]{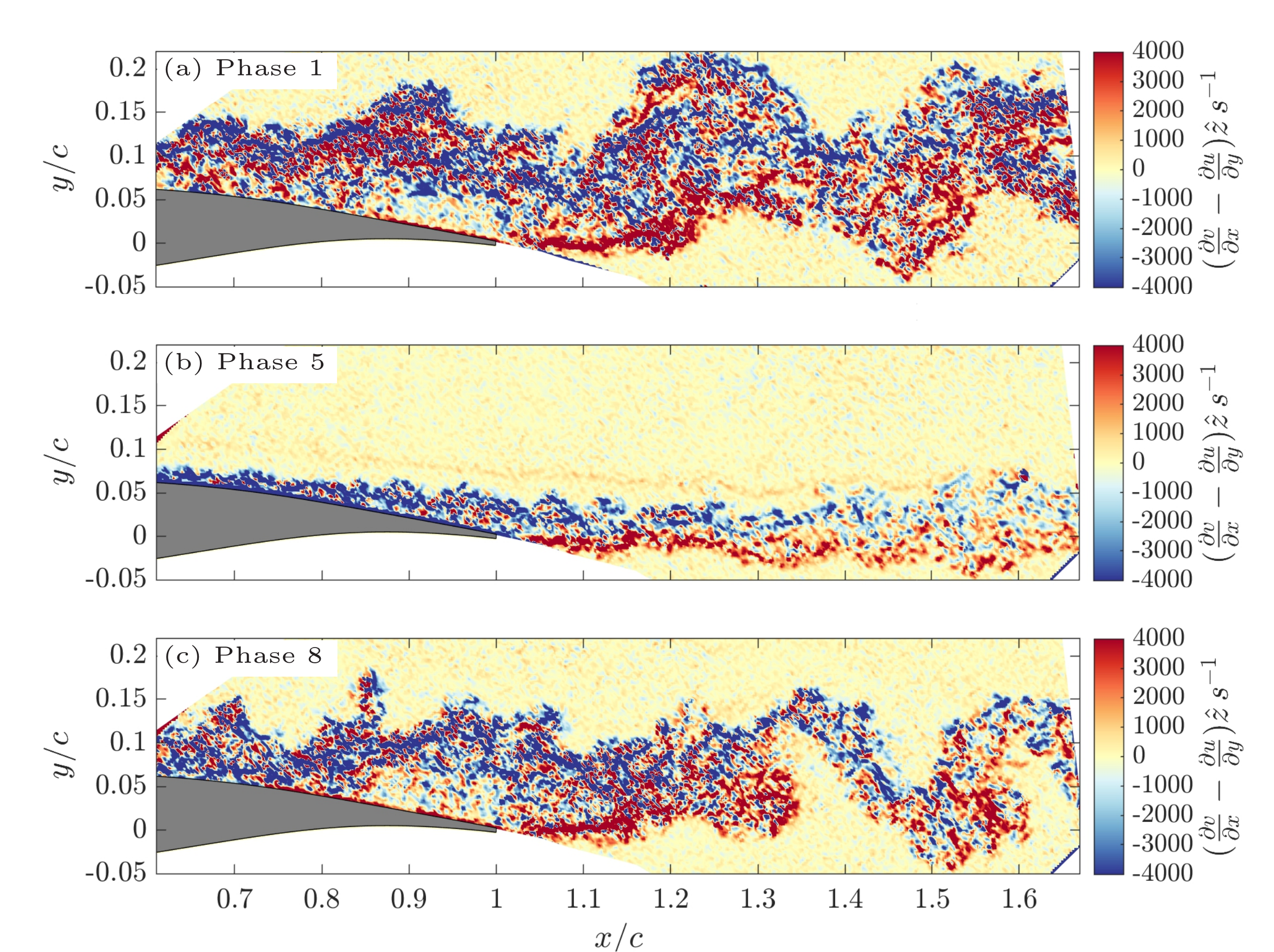}
\caption{Representative instantaneous vorticity contour maps for shock most upstream (a) and downstream (b) locations, and final upstream motion (c).}
\label{fig:Vorticity_Phases_1_5_8}
\end{figure}

%Taking a look at 
Fig. \ref{fig:Vorticity_Phases_1_5_8} %which 
displays the instantaneous vorticity distribution for three characteristic phases of the buffet period: for the most upstream shock location (a), the most downstream shock position (b), and the final phase of upstream motion (c). We observe %, we can clearly study 
the production of vorticity in the near wake and throughout the vortex shedding process at the trailing edge, as well as the strong variation of the regions with intense vorticity %its prominent variation 
coupled with the buffet phase. 

The slip line originating from the triple point of the lambda shock structure is well-resolved by the in-plane vorticity maps derived from PIV: it is visible as a %, and is detected as 
thin streak of moderately positive vorticity that follows the vertical extent of the wake and is almost parallel to the edges of the turbulent wake (see e.g. Fig.\,\ref{fig:Vorticity_Phases_1_5_8} (b)).

%As a global observation, 
We observe coherent vortical structures %are resolved 
across all phases of the buffet cycle, %period, comprising 
and they follow the expected cyclic variation. The contribution from the lower half of the wake is %, but with 
consistently %overall 
positive, and negative vorticity is shed along the upper half.  This tendency was previously observed %captured in a similar fashion 
by Szubert et al. \cite{Szubert2015} based on consecutive snapshots covering one %entire 
vortex shedding period. 

As %was shown 
discussed above, buffet phase 1 %, which 
captures the %flow situation pertaining to 
most upstream shock locations and is associated with massive flow separation with a vertical extent of the turbulent wake of almost $0.2\,c$. The corresponding vorticity map in Fig.  \ref{fig:Vorticity_Phases_1_5_8} (a) reveals two %remarkable details 
aspects of its macroscopic organization: first, Kelvin-Helmholtz-type instabilities develop downstream of the shock wave up to about $x/c=1.2$; second, a pronounced vortex roll-up is observed around the trailing edge and behind. %Taking into account the 
We used an identical scaling of the colorbar as %was used by 
Szubert et al. \cite{Szubert2015} and observe almost perfect agreement with their %present test 
case in terms of vorticity distribution and magnitude. %is recognized.

%Despite qualitative similarities in the observations, t
The quantitative %distinctions 
differences between buffet phases are quite severe. In contrast to the widely separated shear layer and %consequently 
large coherent vortical structures %emanating from 
in the wake of phase 1, phase 5 exhibits small-scale vortical structures confined to a narrow band. The vortical structures protrude less far into the vertical direction, which results %resulting 
in an almost sharp %ly defined 
separation between positive contributions from the lower side of the wake and negative contributions from the upper side (compare Figs.\,\ref{fig:Vorticity_Phases_1_5_8} (a) and (b)). This observation is in good agreement with the findings of %conclusion advanced by 
Epstein et al.\,\cite{Epstein1988}, that an unseparated transonic flow past an airfoil exhibits considerably reduced vortex shedding. %tendency.

The massive thickening of the boundary layer in the aft section of the airfoil results in a pronounced de-cambering effect. Hence, the periodic variation between fully-separated and %widely 
mostly reattached flow %characteristic of the phenomenon 
throughout a buffet cycle results in a continuous alteration of the integral circulation around the airfoil and therefore in continuous %results in permanent 
shedding of vorticity aft of the trailing edge. This effect presumably also contributes to the %repetetively 
strongly alternating %increasing and decreasing 
vortex-shedding intensity.

\subsection{Spectral analysis of the vortex shedding} %in buffet condition}

\begin{figure}[hbt!]
\centering
\includegraphics[width=.98\textwidth]{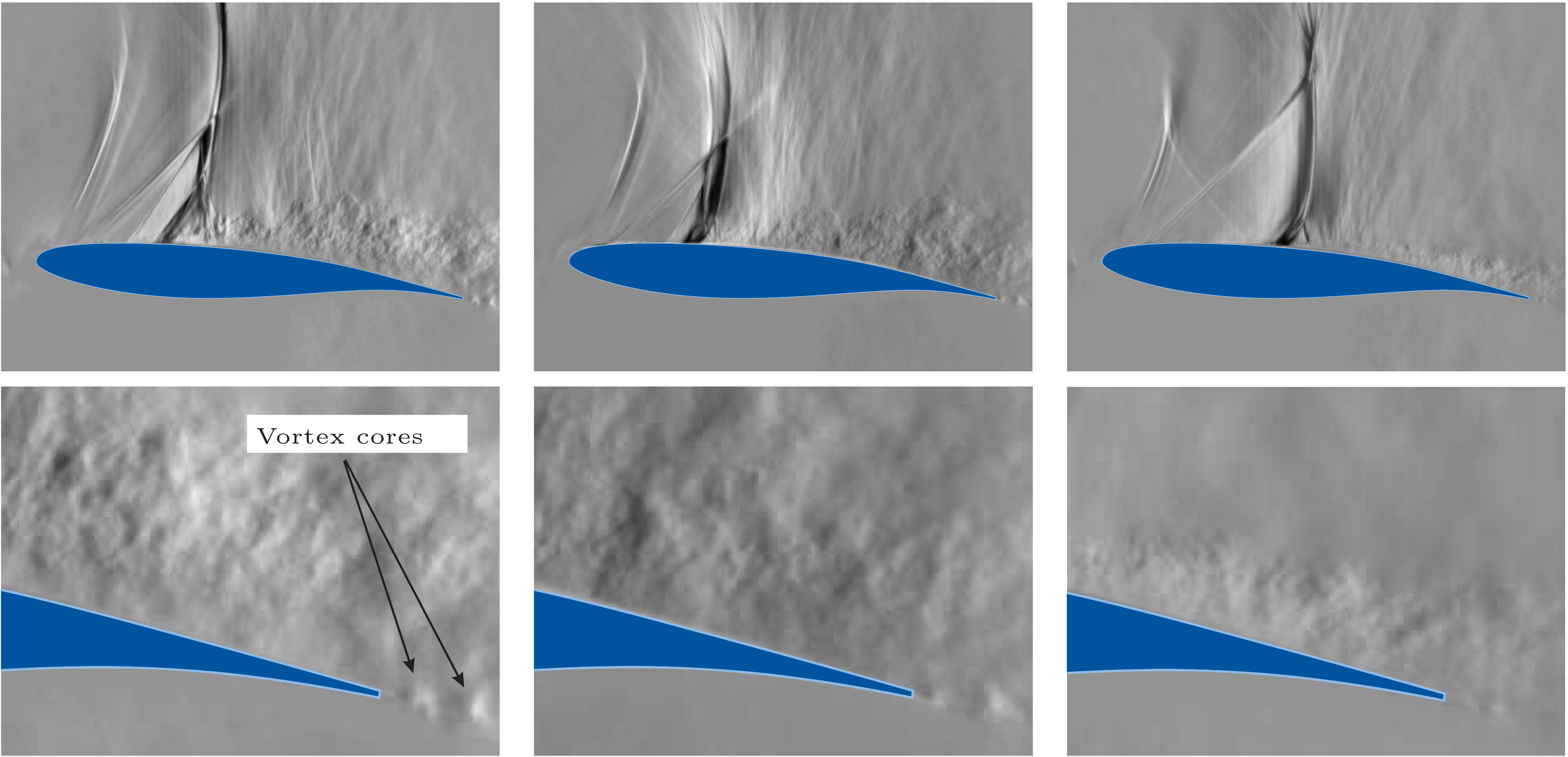}
\caption{%Formation of von Kármán-type vortices 
Schlieren visualizations of the vortex shedding downstream of the trailing edge for shock locations: (a) most upstream, %locations (a), 
(b) incipient downstream motion, and (c) most downstream shock position. Second row: zooms of the trailing-edge region; vortex shedding indicated with arrows.}
\label{fig:SchlierenVortexStructures}
\end{figure}

The vortex cores of the von Kármán vortex street induced upon the interaction of the flow with the trailing edge are resolved by alternating dark and bright spots in the Schlieren visualizations shown in Fig. \ref{fig:SchlierenVortexStructures}. By tracking the time history of the dark/bright transition, we analyze the spectral content of the vortex shedding %process is analyzed 
quantitatively. As %can be seen in the details 
visible in the detail views at the trailing edge (lower row of Fig. \ref{fig:SchlierenVortexStructures}), both the distinctness and spatial extent of the resolved vortex cores decrease from %left to right
phases 1 -- 8 (Fig.\,\ref{fig:SchlierenVortexStructures} (a)-(c)). Assuming that the density gradient %to 
scales with the vortex strength, %we can record that 
the %latter 
strength decreases %progressively 
with increasing chordwise shock location. This observation is in good agreement with the shrinking of the vortex street (see Fig. \ref{fig:Vorticity_Phases_1_5_8}).

%In their study 
Szubert et al. \cite{Szubert2015} pointed out that events of massive flow separation – not necessarily driven by %developed 
transonic buffet – may %give rise to 
induce von Kármán-like vortex shedding in the vicinity of the trailing edge and %beyond in 
the near wake. With pressure sensors placed along the airfoil upper surface and in the near wake, they captured %were able to capture 
strong secondary fluctuations (in addition to the primary buffet oscillation) over a broad range of chordwise stations. These fluctuations -- %in large part 
attributed to a von Kármán mode -- %were shown to be maximal 
reach their maximum at $x/c=1.20$, remain pronounced until %a chordwise station 
$x/c=1.5$, and only decay %substantially 
beyond one chord length downstream of the trailing edge \cite{Szubert2015}. %Based on power spectral density analyses performed on 
In the wall-pressure spectra, Szubert et al. \cite{Szubert2015} observed distinct contributions at higher frequency ($St_{c}=2.5$) close to the trailing edge in addition to the characteristic low-frequency ($St_{c}=0.075$) footprint of shock buffet. %\ams{in the} wall-pressure signals. % showed . 
%With gradually decreasing distance from 
%\ams{Close to} the trailing edge, however, distinct contributions at higher frequency ($St_{c}=2.5$) emerge \cite{Szubert2015}. 
%In perfect agreement with the time history of pressure probes, the corresponding peak in the spectral diagram is greatest at $x/c=1.20$. 
Szubert et al. \cite{Szubert2015} verified the relation of the  $St_{c}=2.5$ peak with the vortex shedding event on the basis of series of instantaneous vorticity maps with adequate temporal resolution. %were subsequently used to track the vortex shedding event, and thus verify its relation with the previously identified frequency peak at $f_{vk}=2600$ or $St_{c}=2.5$.

\begin{figure}[hbt!]
\centering
\includegraphics[width=.7\textwidth]{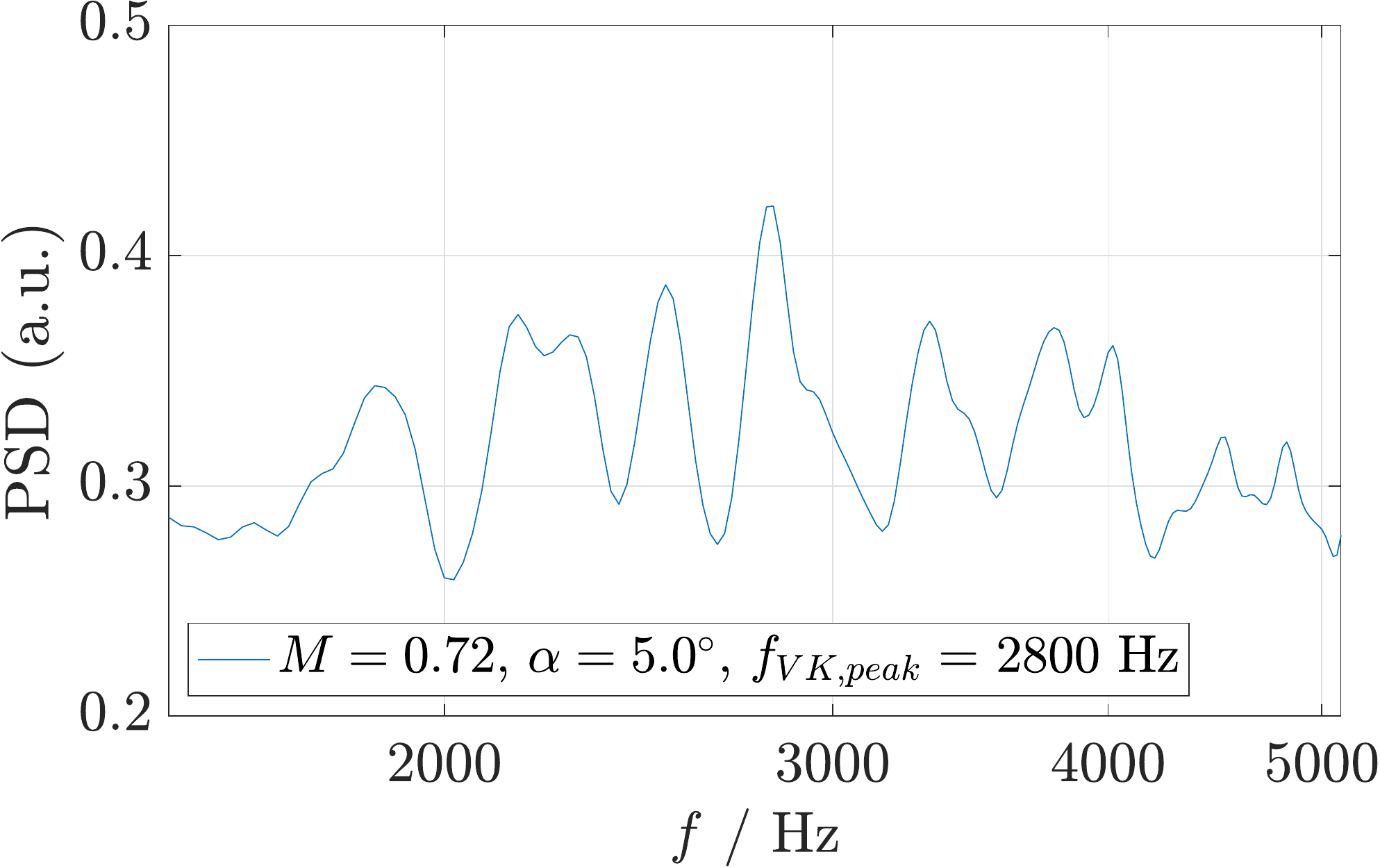}
\caption{Power spectral density distribution of the von Kármán vortex shedding at the trailing edge.}
\label{fig:WelchSpectrumWakeVortex}
\end{figure}

Applying a similar analysis to the present data base yields the spectrum shown in Fig. \ref{fig:WelchSpectrumWakeVortex}. We observe high-frequency contributions peaking at $2800\,\textrm{Hz}$, but with overall enhanced intensity between $2000\,\textrm{Hz}$ and $4000\,\textrm{Hz}$. %In addition to the 
The peak frequency corresponds to $St_c=1.8$, and is thus %being on 
of the same order of magnitude as the study reported above \cite{Szubert2015}.
We also observe a dominant buffet peak at $f_b=113 \, \textrm{Hz}$, or $St_c=0.071$ (not displayed here %for clarity
to improve the visibility of the higher-frequency contributions). %\ams{In addition, we observe} %, noticeable 
%contributions %peaking 
%at $2800\,\textrm{Hz}$, but with overall enhanced intensity between $2000\,\textrm{Hz}$ and $3000\,\textrm{Hz}$ are extracted. The peak frequency corresponds to $St_c=1.8$, thus being on the same order of magnitude as the study reported above \cite{Szubert2015}. 
%Taking into account 
Considering the otherwise flat power spectrum, the pronounced bump can thus be attributed to the vortex shedding process and the incipient formation of a von Kármán vortex street. 

Also Zauner and Sandham \cite{Zauner2020} showed similar alternating large-scale structures (as shown in the present study in instantaneous v velocity (see Fig. \ref{fig:u_phases_instantaneous}) and vorticity plots (see Fig. \ref{fig:Vorticity_Phases_1_5_8})) %were resolved by Zauner and Sandham \cite{Zauner2020} 
based on DMD modes representing both density and streamwise velocity in the wake field between $x/c=0.8$ and $x/c=2.0$. The associated dynamics were centered around $St_c=1.8$, %in that study which are on the same order of magnitude 
which is a similar range as observed in the present case ($St_c=1.3$ to $St_c=1.7$). This agreement is satisfactory, considering %Taking into account 
the  slightly higher Strouhal numbers reported for the overall airfoil configuration in \cite{Zauner2020}, %therein which also comprises 
which also includes slightly higher values %of the buffet phenomenon 
for the buffet-related peak (St=0.12 vs. St=0.07)). %, the results agree well.

\section{Conclusions}\label{sec:Conc}
Transonic buffet strongly affects the aerodynamic performance of passenger aircraft. Due to large-scale shock oscillations, the flow on the suction side of the airfoil is subjected to large-scale intermittent flow separation. Via the wake of the airfoil or wing, also aerodynamic components downstream of the wing, e.g. the tail plane, are strongly affected by this intermittent flow field. To shed light on the flow structure and characteristics impinging on the tail plane, we analyzed the effects of the buffet cycle on the wake in detail.

The buffet-dominated flow fields %in the wall-bound 
along the OAT15A-airfoil suction side %domain 
and in the %near-
airfoil wake were presented and discussed in a complementary way to elucidate the coupling between the shock-induced unsteadiness, the disturbances incited at the trailing edge, and the vortical structures in the wake. %It was demonstrated that t
The %buffet-dominated 
flow field on the airfoil suction side undergoes a strong periodic and global variation %induced by 
throughout the buffet cycle, and also the wake (shown until $x/c=1.7$) is dominated by %that also strictly involves major parts of the near wake domain up to $x/c=1.7$ showing evidence of 
pronounced periodicity. Owing to the large-scale shock displacement coupled with an intermittent separation, the flow topology both along the %wall-bound domain 
airfoil and in the near wake is modified on a global scale. The associated fluctuations remain at a constantly high level throughout the entire studied domain ($x/c=1.7$) %up to about $70\,\%$ downstream of the airfoil 
and are suspected to %exist even further 
persist also farther downstream. 

On the basis of phase-averaged streamwise velocity maps, we showed the flow's periodic %revealed a cyclic 
variation, which is synchronized with the buffet cycle: the low-velocity wake core fluctuates strongly, %with a strong fluctuation of , mostly 
dictated by the respective location and direction of motion of the shock wave. The wake is %predominantly defined 
dominated by a pronounced variation of the vertical and streamwise extent of the velocity deficit associated with the %intense 
flapping motion of the separated shear layer. Each buffet phase has a characteristic footprint of the flow topology, %associated with  %could be derived that is most strikingly 
which is defined by the turbulent intensity %of turbulent fluctuations 
and %an 
effective angular evolution of the main wake axis. 

Downstream of the trailing edge, vortical structures form and %that 
are organized as large coherent %vortical 
streaks in phases where the wake is largest. They decrease %, and reduce 
substantially for phases of reattached flow. %where the flow is} mostly reattached. %flow situations. One observes that p
Positive vortical motion emanates at the lower edge of the wake and is intertwined with negative vorticity from the upper region. Consequently, %thus forming 
Kármán vortex-street-like configurations form. 

The low-frequency buffet unsteadiness %is identified to 
modulates the vortex shedding, as it drastically modifies the circulation around and %the flow 
past the airfoil. Spectral analyses of the vortex shedding revealed that the %involved 
associated frequencies are one order of magnitude larger than those of the buffet mode ($St=\mathcal{O}({1})$) and $St=\mathcal{O}({0.1})$, respectively). 
These structures and associated %characteristic 
frequencies will affect any downstream aerodynamic devices, such as the tail plane, on which the wake of the airfoil impinges. It is therefore to expect that the flow around the tail plane is also strongly periodic, with %strongly 
varying turbulent intensities associated with distinctly dominating peak frequencies  in the range of the shock-buffet %frequency 
($St_c=0.07$) and %the 
vortex shedding frequencies ($St_c=1.8$).

\section*{Acknowldgements}

The authors gratefully acknowledge the German Research Foundation (Deutsche Forschungsgemeinschaft DFG) for funding this work in the framework of the research unit FOR 2895 (project number 406435057). The authors %wish to 
thank ONERA for providing the OAT15A airfoil geometry for our %that was used to manufacture the 
wind tunnel model. We also gratefully acknowledge the contribution and support of Nick Capellmann and the %entire 
workshop team of the Institute of Aerodynamics during the manufacturing process of the wind tunnel model.

%Some journals require declarations to be submitted in a standardised format. Please check the Instructions for Authors of the journal to which you are submitting to see if you need to complete this section. If yes, your manuscript must contain the following sections under the heading `Declarations':
%
%\begin{itemize}
%\item Conflict of interest/Competing interests (check journal-specific guidelines for which heading to use)
%\item Ethics approval 
%\item Consent for publication
%\item Availability of data and materials
%\item Authors' contributions
%\end{itemize}
%
%\noindent
%If any of the sections are not relevant to your manuscript, please include the heading and write `Not applicable' for that section. 




\end{document}